\documentclass[twocolumn,showpacs,preprintnumbers,amsmath,amssymb,superscriptaddress,nofootinbib,groupedaddress,floatfix]{revtex4-1}
\usepackage{graphicx} % needed for figures
\usepackage{dcolumn} % needed for some tables
\usepackage{bm} % for math
\usepackage{amssymb} % for math
\usepackage{amsmath} % for align environment
\usepackage{amsfonts}
\usepackage{color}
\usepackage{comment}
\usepackage{rotating}
\usepackage{multirow}
\usepackage{ctable}
\usepackage{colortbl}
\usepackage{multirow}
\usepackage{afterpage}
\usepackage{subcaption}
\usepackage{hyperref}
\usepackage{soul}
\usepackage{array}

\usepackage[table]{xcolor}
\definecolor{coneYellow}{RGB}{255,240,0}

\usepackage{array}
\newcommand{\PreserveBackslash}[1]{\let\temp=\\#1\let\\=\temp}
\newcolumntype{C}[1]{>{\PreserveBackslash\centering}p{#1}}
\newcolumntype{R}[1]{>{\PreserveBackslash\raggedleft}p{#1}}
\newcolumntype{L}[1]{>{\PreserveBackslash\raggedright}p{#1}}

\definecolor{darkgreen}{RGB}{50,150,0}

\begin{document}

\title{Evolving Dark Sector and the Dark Dimension Scenario}
\author{Alek Bedroya$^{1}$, Georges Obied$^{2}$, Cumrun Vafa$^{3}$, David H. Wu$^{3}$}
\affiliation{\small ${}^{1}$
{\it Princeton Gravity Initiative, Princeton University, Princeton, NJ 08544, USA} \\
\small ${}^{2}$
{\it Rudolf Peierls Centre for Theoretical Physics, University of Oxford, Oxford, OX1 3PU,
United Kingdom}\\
\small ${}^{3}$
{\it Jefferson Physical Laboratory, Harvard University, Cambridge, MA 02138, USA} }

\begin{abstract}
  String theory naturally leads to the expectation that dark energy is not stable, and may be evolving as captured by the Swampland de Sitter conjectures.  Moreover, motivated by the distance conjecture, a unification of dark sector has been proposed, where the smallness of dark energy leads to one extra dimension of micron size with dark matter being the Kaluza--Klein graviton excitations in this extra dimension.  We consider the natural possibility that the radius of the dark dimension varies as the energy decreases, leading to the variation of the dark matter mass.  This correlates the variation of the dark energy with the variation of the dark matter mass as they depend on the variations of a scalar field $\phi$ controlling the radius of the extra dimension.  A simple realization of this idea for small range of $\phi$ is adequately captured by choosing a potential which is locally of the form $V=V_0\ {\rm exp}(-c\phi)$ and dark matter mass $m_{\rm DM}=m_0\ {\rm exp}(-c' \phi)$ as was proposed in~\cite{Agrawal:2019dlm}. We find excellent agreement with recent experimental data from DESI DR2 combined with SN measurements (from DES, Union3 or Pantheon+) and reproduces the same significance as CPL parametrization with the added benefit of providing a natural explanation for the apparent phantom behavior ($w<-1$) reported by DESI and DES based on a physical model. DESI and SN datasets independently favor non-zero values of \( c' \) and \( c \), respectively, both lying within the expected \(\mathcal{O}(1)\) range suggested by the Swampland criteria.  Moreover, our best fit value $c'\simeq 0.05 \pm 0.01$ is remarkably consistent with the experimental upper bound of $c'\lesssim 0.2$ demanded by the lack of detection of fifth force in the dark sector. 
\end{abstract}

\maketitle
\section{Introduction}

In non-gravitational theories, a common paradigm that guides model building is the concept of distance scale separation, which is the idea that microscopic physics has limited influence on macroscopic phenomena. This leads to the idea that for an EFT in the infrared (IR) one can ignore the ultraviolet (UV) description as all the allowed terms in the action are subject to symmetries of the theory where the strengths of the higher derivative terms are set by the UV cutoff. This principle underlies many cosmological models that aim to describe the universe on the largest scales. However, in gravitational theories, it has been understood that UV and IR physics are deeply interconnected. The \textit{Swampland program} 
(see \cite{Agmon:2022thq,vanBeest:2021lhn,Palti:2019pca} for reviews), rooted in lessons learned from string theory, seeks to make the IR implications of a consistent gravitational UV completion more precise. In particular, the form of the scalar potential and the spectrum of light states---both crucial to the cosmological history of the universe---are tightly constrained by the requirement of UV completeness in quantum gravity. 

The aim of this letter is to revisit the impact of these constraints on the recent cosmological epoch.  We undertake this in order to assess the status of recent observations and their compatibility with these constraints. We are particularly interested in the emerging evidence for varying dark energy density. There were hints of this behavior in the DES Y5 data release~\cite{DES:2024jxu} and this has since increased in significance in light of the new DESI data ~\cite{DESI:2025zgx,DESI:2025fii}. In this context, we find a strikingly good match with the Swampland constraints!  There are two facets to these restrictions:  On the one hand, based on the \textit{Distance Conjecture}~\cite{Ooguri:2006in} and the related conjectures of the \textit{AdS Distance Conjecture}~\cite{Lust:2019zwm} and the \textit{Emergent String Conjecture}~\cite{Lee:2019wij}, extreme values of physical parameters lead to the expectation that we have either large extra dimensions with a light Kaluza--Klein (KK) tower or a fundamental string with a light tower of string excitations. Applied to the smallness of dark energy, this has led to a unification of the dark sector through the prediction of one extra dimension in the micron range, and the light KK graviton tower as constituting the dark matter (the \textit{Dark Dimension Scenario}~\cite{Montero:2022prj,Gonzalo:2022jac}).  On the other hand, positive values of potential $V$ are expected to be unstable in a particular way  (the \textit{de Sitter Conjecture}~\cite{Obied:2018sgi} and the \textit{Trans-Planckian Censorship Conjecture (TCC)}~\cite{Bedroya:2019snp}). See~\cite{Vafa:2024fpx,Vafa:2025nst} for a basic review of the applications of these principles to phenomenology.

In \cite{Agrawal:2018own}, it was shown that the dS conjecture is compatible with cosmological observations. 
 Moreover, motivated by these conjectures, Swampland compatible models of evolving dark energy have been considered in many works. In particular, in \cite{Agrawal:2019dlm} it was pointed out that the natural unification of the dark energy and dark sector as implied by the Swampland criteria would lead to the fading of the dark matter (decreasing of $\rho_M a^3)$ as the universe evolves and the dark energy depletes.  The main aim of this paper is to reevaluate this proposal in light of the dark dimension scenario and the recent DES and DESI DR2 results.  In particular, we view the evolving length of the dark dimension as the mechanism that lowers the effective energy (including the matter contribution).  Indeed, we find that this scenario is in excellent agreement with data, with preference for expansion of dark dimension in matter dominated era and shrinking of dark dimension in recent epoch which reproduces the same significance as the CPL parametrization with the benefit of providing a natural explanation for the apparent phantom behavior ($w < -1$) reported by DESI and DES.

The organization of this paper is as follows.  In Section~\ref{sec:2}, we review the relevant Swampland conjectures.  In Section~\ref{sec:3}, we discuss how these lead to the dark dimension scenario and motivate an evolving dark sector as the radius of the dark dimension evolves.  Finally, in Section~\ref{sec:4}, we assess the compatibility of these predictions with the DESI DR2 results in combination with various other datasets.

\section{Dark sector and the Swampland conjectures}\label{sec:2}

\subsection{Scalar potential}

In gravitational theories, the scalar potential backreacts on the spacetime metric unlike non-gravitational theories where it can be shifted by a constant without physical consequence. Therefore, it is not surprising to expect the behavior of the scalar potential to be constrained in gravitational theories. 

In all known weak-coupling limits of the string landscape where the scalar potential can be perturbatively calculated, it always falls off exponentially fast in terms of the canonical distance in the scalar field space. Moreover, the fall-off rate appears to be bounded from below. These observations were formalized in the \textbf{de Sitter Conjecture (dSC)}, which postulates that in any consistent theory of quantum gravity, the scalar potential at any point in field space satisfies one of the following two inequalities for fixed positive numbers $c_1$ and $c_2$ of $O(1)$ in Planck units~\cite{Obied:2018sgi,Ooguri:2018wrx,Garg:2018reu} (throughout this paper if not explicitly stated we work in Planck units, where reduced $M_{\rm pl}=1$)
\begin{align}\label{dsc}
    |\nabla V| > c_1 V \quad \text{or} \quad \{{\nabla_i\nabla_j} V\} < -c_2 V\,.
\end{align}
The norm of the variation $\nabla V$ of the scalar potential is defined with respect to the canonical metric on scalar field space determined by the kinetic term. Moreover, $\{{\nabla_i\nabla_j} V\}$ denotes the smallest eigenvalue of the Hessian matrix of $V$ with respect to the same metric.

The first condition in \eqref{dsc} implies that positive scalar potentials must be sufficiently steep, while the second allows for exceptions only when the potential is sufficiently concave, such as at a local maximum. In particular, this conjecture rules out the existence of meta-stable de Sitter vacua realized as local minima of the scalar potential. Despite extensive efforts to construct meta-stable de Sitter vacua in the string landscape, no fully controlled example has yet been achieved. However, a recent construction based on symmetries in string theory has realized an unstable de Sitter space consistent with the dSC~\cite{Chen:2025rkb}.

A related but distinct conjecture that makes the dSC quantitatively predictive is the \textit{Trans-Planckian Censorship Conjecture (TCC)}~\cite{Bedroya:2019snp}, which recasts dSC in terms of a physical principle. TCC postulates that any period of accelerated expansion must be limited so that Planckian fluctuations do not exit the Hubble horizon. More precisely, an FRW expanding universe with initial scale factor $a_i$ that evolves to a final scale factor $a_f$ and Hubble parameter $H_f$ must satisfy
\begin{align}
     \frac{a_f}{a_i}  M_{\rm pl}^{-1}<H_f^{-1}\,.
\end{align}

This condition is a clear example of the deep connection between UV and IR physics in gravitational theories. Applying the TCC to exponential potentials in $d$-dimensional spacetimes implies that their fall-off rate in the infinite distance limits of field space must satisfy (in Planck units):
\begin{align}\label{dsb}
   V(\phi) \sim \exp(-\lambda \phi) \quad \Rightarrow \quad \lambda \geq \frac{2}{\sqrt{d-2}}=\sqrt{2}\quad {\it for}\ d=4\,.
\end{align}
TCC also implies a weaker bound in the interior of field space. Assuming the potential decreases monotonically away from a local maximum at $\phi = 0$, TCC suggests that  
\begin{align}\label{intTCC}
    V(\phi) &\lesssim \exp\left(-\frac{2|\phi|}{\sqrt{(d-1)(d-2)}}\right)\nonumber\\
    &\Rightarrow \lambda_{\rm interior}\geq \sqrt\frac{2}{3}\approx 0.82\quad {\it for}\ d=4\,.
\end{align}
The above bound is expected to hold in an intermediate regime in the interior of the scalar field space that interpolates between the local maximum and the infinite-distance limits and implies that the mean value of $|\nabla V|/V$ in the interior satisfies this bound. Remarkably, the same inequality was also independently derived in~\cite{vandeHeisteeg:2023uxj} using the Emergent String Conjecture. 

The bound (\ref{dsb}) is consistent with string theory examples (see \cite{Andriot:2022xjh}) and also invariant under dimensional reduction~\cite{Rudelius:2021azq}, and has been independently motivated by holographic arguments in infinite distance limits of field space~\cite{Bedroya:2022tbh}.
In certain cases, the TCC follows from other Swampland conjectures~\cite{vandeHeisteeg:2023uxj,Bedroya:2025ris}, and can even be argued to hold in the interior of moduli space for specific classes of potentials~\cite{Bedroya:2024zta,Bedroya:2025ris,vandeHeisteeg:2023uxj}.

As pointed out in~\cite{Bedroya:2019snp}, a constant dark energy is compatible with the TCC as long as it decays not too long after Hubble time.  This conjecture essentially implies that the duration of any quasi-de Sitter expansion must be finite, whether it is driven by a rolling scalar field or realized as a metastable dS vacuum (leading to lifetime $\tau <(1/H) {\rm log}(1/H)$.  Interestingly this upper bound on lifetime is also the same conclusion one draws from dSC for a hilltop potential, where the maximum lifetime is no larger than $\tau\lesssim  {\rm log}(1/H)/H$ where $H^2\sim V_{max}$. However, the dSC and TCC imply that a more natural alternative capable of sustaining classical expansion for a number of e-folds is one driven by a scalar field potential whose steepness is proportional to the potential itself
\begin{align}
    |\nabla V|\sim V\,.
\end{align}
In particular, this implies that a small value of the scalar potential is naturally accompanied by a small value of its derivative, thereby linking the fine-tuning problems associated with the smallness of the slope with the smallness of its value.
It is natural to expect that $V$ which asymptotically approaches an exponential potential with a relative slope $|\nabla V|/V$ larger than $\sqrt{2}$, has a smaller relative slope as we move away from the asymptotic limit (otherwise the potential will become larger than the Planck scale). In particular, in the interior of the field space, we therefore expect the slope to decrease, leading to a potential of the form $V \sim \exp(-c\Delta \phi)$ where $c$ may be smaller than $\sqrt{2}$. Such a scalar potential of $\rm{sech}$ type with asymptotic slope $\sqrt{2} $ was shown to be consistent with recent DESI DR2 results~\cite{Anchordoqui:2025fgz} (see also \cite{Brandenberger:2025hof,Andriot:2025los,Anchordoqui:2025epz} for additional work studying the recent data from the perspective of Swampland principles).

\subsection{Tower of light states}

Another ubiquitous feature in the quantum gravity landscape is the presence of dualities, which relate different weakly-coupled descriptions that share the same scalar field space. Moreover, the weakly-coupled descriptions emerging in the infinite-distance limits of the scalar field space always involve a tower of light states whose masses have a specific dependence on the scalar fields. This observation was formalized in the statement of the \textit{Distance/Duality Conjecture (DC)} \cite{Ooguri:2006in}, which postulates that when one takes the value of a scalar field to an infinite-distance limit, there is always a tower of light states whose mass exponentially decays in terms of the canonical distance defined by the metric on the moduli space.

Furthermore, based on a wide range of constructions and observations in string theory, this conjecture was refined into the \textit{Emergent String Conjecture (ESC)}, which postulates that the tower of light states are always either a tower of Kaluza--Klein states descending from a higher-dimensional theory or are excitations of a fundamental string which includes the graviton as a string state~\cite{Lee:2019wij}.

As shown in~\cite{Agmon:2022thq}, this conjecture has quantitative implications, such as the sharpened Distance Conjecture~\cite{Etheredge:2022opl}.  In particular, in infinite distance directions we have a tower with a mass scale $m\sim {\rm exp}(-\lambda' \phi)$ such that
\begin{align}\label{dcb}
    \sqrt{\frac{d-1}{d-2}}\geq \lambda'  \geq \frac{1}{\sqrt{d-2}}\Rightarrow \sqrt{\frac{3}{2}}\geq \lambda'\geq \frac{1}{\sqrt 2}\quad {\it for}\ d=4\,.
\end{align}
In particular, by combining \eqref{dsb} and \eqref{dcb}, we learn asymptotically in field space in 4d
$$\lambda \geq \sqrt{\frac{4}{3}}\lambda'\,.$$
However, one can in fact argue for a stronger bound asymptotically~\cite{Andriot:2020lea}:

\begin{align}\label{srb}
    \Bigg|\frac{\nabla V}{V}\Bigg|\geq 2\Bigg|\frac{\nabla m}{m}\Bigg|\,.
\end{align}
This follows from the requirement that $V$ must remain smaller than $m^2$, and therefore fall off sufficiently fast. In quasi-de Sitter spaces, the inequality $V = 3\Lambda^2 \leq \frac{3}{2}m^2$ follows from the Higuchi bound~\cite{Higuchi:1986py} which bounds the mass of massive spin-2 particles based on unitarity and can be applied to the particles in the tower which would lead to the expectation of the above bound (not exactly because the bound refers to slopes).
However, as we move away from quasi-de Sitter expansion, the Higuchi bound no longer applies. Nevertheless, the inequality~\eqref{srb} can still be argued for. We divide the argument into two cases. First, if $|\nabla V/V| > 2\sqrt{d-1}/\sqrt{d-2}$, it follows directly from~\eqref{dcb} that~\eqref{srb} is satisfied. Second, if $|\nabla V/V| \leq 2\sqrt{d-1}/\sqrt{d-2}$, the scalar potential falls off sufficiently slowly to drive the late-time cosmological evolution in a spatially flat FRW solution (see Appendix A of~\cite{Bedroya:2022tbh}), implying $V \propto H^2$. For open universes, $V \propto H^2$ at late times irrespective of the slope of $V$~\cite{Andriot:2023wvg,Bedroya:2025ris}. Since the smallest $d$-dimensional black hole has a radius of order $m^{-1}$~\cite{Bedroya:2024uva}, requiring that a black hole can fit within the observable universe imposes $H < m$. This, in turn, leads to $V \lesssim m^2$ at any point in the moduli space, as also argued in~\cite{Bedroya:2025ris}, thereby leading to~\eqref{srb}.

The distance conjecture has also been applied to the case when the cosmological constant is small~\cite{Lust:2019zwm}. In this case, unlike the discussion for the scalar field $\phi$ above, the distance refers to movement in flux space and other parameters of the landscape which leads to a small cosmological constant. In particular, it has been argued that in this limit, there is a tower of light states whose mass scale, {\it rather than its slope in field space}, correlates with the value for $V=\Lambda$:
$$m\sim V^{\alpha}\,,$$ 
where $\alpha\sim O(1)$.
Moreover, it was argued in~\cite{Montero:2022prj} that $\frac{1}{d}\leq \alpha \leq \frac{1}{2}$ (see also the recent work \cite{Herraez:2025gcf}), or equivalently 
\begin{align}\label{DDC}
V\sim m^{\beta},\quad 2\leq \beta \leq 4\quad {\rm for}\ d=4.
\end{align} 
To avoid confusion, it is important to note that we have two distinct types of relations between $V$ and $m$ referring to asymptotics in flux vacua vs. in scalar field space.  The first case relates the {\it value} of $V$ and $m$.  The second relates the {\it slope} of $V$ in field space, to that of $m$.  For example, in the context of AdS/CFT, if you consider the ${\cal N}=4$ $SU(N)$ SYM in $d=4$, we can consider the limit $N\gg 1$ but fixed coupling $\tau\sim \tau_0$.  In such a case, we end up with $|\Lambda|\sim N^{-4/3}$, with mass tower $m\sim |\Lambda|^{1/2}$, but we may not be in the asymptotic limit of $\tau$ (e.g., we could be at $\tau =i$). As we will discuss in the next section, for our universe, we expect to be in the asymptotic limit of flux vacua space, but as the cosmological data shows we are {\it not at the asymptotic limit of scalar field space}.

\section{Dark Dimension Scenario}\label{sec:3}

As discussed in the previous section, if $\Lambda\ll 1$ we expect a tower of light string states or KK gravitons.  It was argued in~\cite{Montero:2022prj} that given the observations and the relation between the tower mass scale $m$ and $\Lambda$, the only option in our universe is to have one mesoscopic extra dimension\footnote{It was recently suggested in \cite{Anchordoqui:2025nmb} that there is a highly fine-tuned scenario with a small window in the parameter space for having two extra dark dimensions with $L<1\mu m$ with 6d Planck scale ${\hat M}_p\sim 10\ {\rm TeV}$.  However, for such a low Planck scale, one would expect to have already seen evidence of new gravitational physics at the LHC. One motivation for considering 2 extra dimensions is the assumption that it may help with electroweak hierachy problem through EFT arguments \cite{Arkani-Hamed:1998jmv}.  However, not only is the 10 TeV scale somewhat higher than the weak scale to explain this, but also the same EFT arguments, unlike the Swampland reasoning, would lead to the incorrect prediction that $\Lambda^{1/4}$ is in the electroweak scale.  Electroweak hierarchy has rather different potential explanations in the Swampland motivated dark dimension scenario \cite{Montero:2022prj,Gonzalo:2022jac}.} with length scale in the micron range $0.1\mu m \lesssim L\lesssim 10 \mu m$.   Moreover, the standard model fields are localized in the extra dimension on a 3+1 dimensional subspace.  The relation between the KK graviton mass scale $m$ and cosmological constant saturates the inequality \eqref{DDC} with $\beta \simeq 4$:
$$\Lambda \sim A \ m^4\,,$$
with $m\sim 1/L$ and for some order 1 parameter $A$, which leads to a micron scale extra dimension.
 Moreover, it was suggested that the associated light tower, the KK graviton in the 5-th dimension, serves as a natural \cite{Gonzalo:2022jac} and viable candidate \cite{Obied:2023clp,Law-Smith:2023czn} for dark matter (see \cite{Vafa:2024fpx} for a review).
 This is a dynamical model for dark matter \cite{Dienes:2011ja,*Dienes:2011sa}.
 In the scenario of~\cite{Gonzalo:2022jac} there is intra-tower KK graviton decays which on its own does not appreciably change the total mass of the dark matter due to an approximate conservation of KK number which follows from the smoothness of the dark dimension.  Note, however, the total mass of the dark matter sector \textit{can} change if the length of the dark dimension changes, as the tower states have masses roughly of the order of $n/L$ for integer spaced quanta $n$.  Indeed, we will consider this possibility in this paper (for another approach to dark matter in dark dimensions see \cite{Anchordoqui:2022svl,Anchordoqui:2022tgp,Anchordoqui:2022txe,Anchordoqui:2024akj}). 

Whether the dark dimension length $L$ is possible to fix or not depends on whether we assume the dS conjecture, or TCC.  Assuming dSC, there cannot exist a metastable length scale since $V>0$.  Assuming TCC instead, if the modulus controlling the size of the extra dimension is not rolling, there could be a rather short-lived meta-stable state (with lifetime $\tau \lesssim (1/H)\log(1/H)$).  In this paper, we would be most interested in the case where the extra dimension is not stabilized and evolves.  In particular, we have a potential $V(L)$ and a tower of states with masses $m\sim 1/L$ both of which evolve.  It is natural to assume that $L$ evolves by growing or shrinking leading to decrease or increase in dark matter mass $m$. 
More generally, we can write both $V$ and $m$ as a function of a scalar field $\phi$ that can change $L$ and possibly other internal moduli of the microscopic dimensions.  In particular, we would be considering both the possibility that the Newton constant is fixed or not in units of mass scales defined using Standard model parameters (e.g., Higgs mass). The former possibility is achieved by taking the total volume of the internal dimension to be constant\footnote{Note that in this case the directions of $\nabla m$ and $\nabla V$ will not align, which would be compatible with observing a smaller slope for $m$ along the direction of $\nabla V$.}, i.e. the change in the dark dimension length is compensated by a change in the volume of the other micrsocopic dimensions. The other case of varying Newton's constant is achieved by a partial variation in the total volume of the extra dimensions. This is also a theoretically valid possibility.  There are interesting observational bounds on the variations of Newton's constant~\cite{Uzan:2024ded} and it provides another opportunity to make contact with experiment if we assume the total internal volume changes.

\begin{figure}
    \centering
    \includegraphics[width=1.1\linewidth]{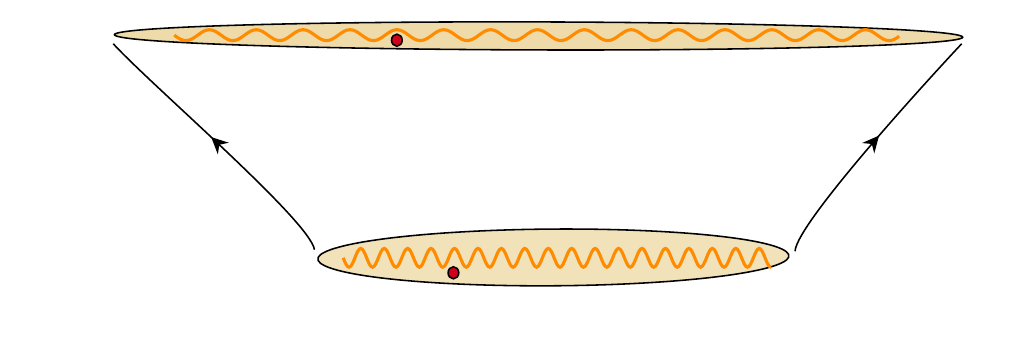}
    \caption{The above figure illustrates the possibility that a large extra dimension is expanding while the overall volume of the internal manifold may or may not remain fixed. Volume can remain fixed through a simultaneous contraction of other internal dimensions, thereby preventing any variation in Newton’s constant. As the fifth dimension expands, the KK gravitons associated with this dimension redshift, leading to a decreasing effective mass from the four-dimensional perspective.  The Standard Model is localized on a brane (depicted as the red dot) that does not wrap any of the evolving dimensions and its moduli can remain naturally frozen.
}
    \label{fig2}
\end{figure}

Note that the parameters of the standard model (SM) would not need to change as they are localized in the dark dimension and the change of the overall length of the dark dimension does not directly impact the local geometry of the standard model brane (Fig.~\ref{fig2}).  Indeed, we will assume that the microscopic moduli of the SM brane are frozen, consistent with strong restrictions on the time variations of the parameters of the SM.  In the next section, we will consider the cosmological model based on this scenario.

\section{Application to Cosmology}\label{sec:4}

From the dark dimension scenario, we are left with two possibilities regarding the dark energy:  if we assume dS conjecture, the dark energy must be decreasing due to a scalar field rolling down the scalar potential; or if instead we assume the TCC,
we can allow either this rolling behavior or a cosmological constant arising from a short-lived metastable de Sitter phase. Given the recent observations of DESI, the data clearly favors the rolling option.    

\begin{figure}
    \centering
    \includegraphics[width=1.05\linewidth]{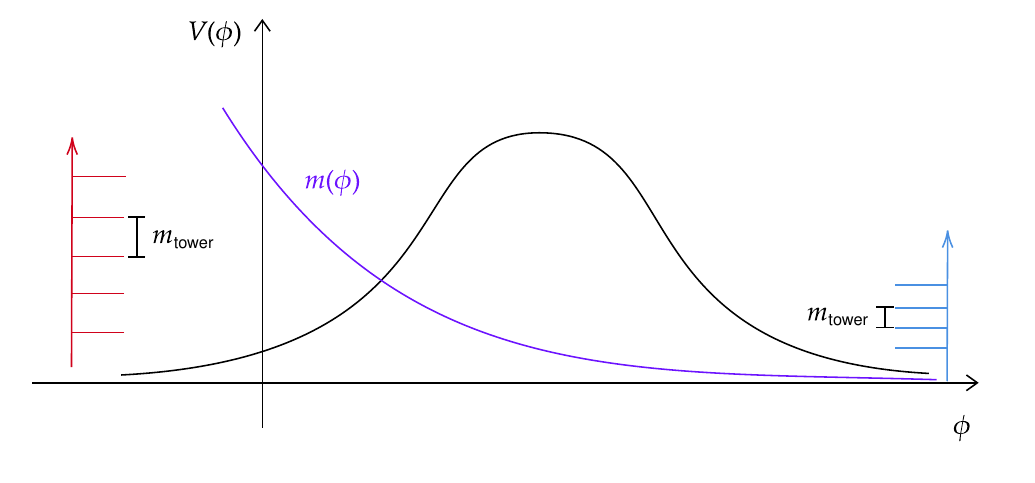}
    \caption{The scalar potential asymptotes to zero at weak-coupling limits, which can arise in opposite scalar field directions. In the dark energy dominated era, a scalar field value to the right (left) of the peak leads to a motion in the positive (negative) $\phi$-direction corresponding to an expanding (contracting) extra dimension. The local gradients $\nabla V$ and $\nabla m$ need not align, leading to $c$ and $c'$ with possibly opposite signs.
}
    \label{fig:fade}
\end{figure}

We know experimentally that we are not quite in the asymptotic limit of $\phi$, because in particular we would have deduced $|V'/V|\geq \sqrt 2$, which would not be consistent with cosmological observations. The question is how to parameterize $V$ and $m$ as a function of the rolling field $\phi$.  For the purposes of modeling this, it turns out since the movement in field space in our universe is small, only the first order variation of the scalar field space really matters.  Given that the logarithms of $V,m$ are the natural parameters, as seen from Swampland arguments, we locally parameterize $V$ and $m$ by
$$V=V_0 \ {\rm exp}(-c \delta \phi)\,,\quad m=m_0\ {\rm exp}(-c'\delta \phi)\,.$$
We define $\phi$ such that the mass decreases in the positive $\phi$ direction which means $c'(\phi)=-\partial_\phi {\log(m)}$ is positive while the sign of $c(\phi)=-\partial_\phi {\log(V)}$ is undetermined. The coefficients $c$ and $c'$ should be viewed as field gradients of $\log(V)$ and $\log(m)$ which may vary with $\phi$.  If the variation of $\phi$ is not large, first-order variation will suffice and we will treat $c,c'$ as constants in this paper, though in principle one can consider power series in $\delta \phi$.
Note that $m$ should be interpreted as the mass scale that sets the spacing between the states in the tower, rather than as the mass of the individual dark matter particles (see Fig.~\ref{fig:fade}). For simplicity of notation, from here on we replace $\delta \phi\rightarrow \phi$, but it should be understood that we simply perform a power series expansion for $c$ and $c'$ and keep the first term.  

It is natural to expect that both $c,c'\sim O(1)$, though they do not need to satisfy the asymptotic bounds which would demand 
$$(c,c')_{\phi \rightarrow \infty}\geq (\sqrt 2,\frac{1}{\sqrt 2})\approx (1.4,0.7)\,.$$
Moreover we expect that the slope for the potential as well as the mass decreases away from the asymptotic limit, especially if we assume $V$ is decreasing in all asymptotic regions in field space.  So in the interior we could easily have $(|c|,c')_{interior}<(1.4,0.7)$.
Similarly, we do not need to satisfy the asymptotic bound for the ratio:
\begin{align}\label{srb1}
    \Big(\frac{|c|}{c'}\Big)_{\phi \rightarrow \infty}\geq 2\,.
\end{align}
But these benchmark values would be interesting to keep in mind as the deviations from them will in a sense mark how far away we are from the asymptotic regime of $\phi$.  As we shall see shortly, the experimental observations yield values for $c,c'$ which are close to $\mathcal{O}(1)$ numbers, which is reassuring.

The directions of $\nabla m$ and $\nabla V$ do not need to align. In the dark dimension scenario, $m$ is only a function of the size of the fifth dimension. Assuming that there is only one light scalar field $\phi$ which is parallel to $\nabla V$, depending on the change in the size of the fifth dimension in that direction, $c'$ can be much smaller than 1 in Planck units\footnote{As seen in Fig.~\ref{fig:FDS 2 scalar fields}, data does not prefer a cosmology driven by a more complicated 2D trajectory in the scalar field space.}. For example, if changing $\phi$ leaves the diameter of the fifth dimension intact, $c'$ would vanish. However, a more natural possibility is that $c'$ is $\mathcal{O}(1)$ in Planck units indicating that dark dimension diameter varies as $\phi$ flows. We choose the direction of $\phi$ such that $m$ decreases as $\phi$ increases. Note that if the inner product $\nabla m \cdot \nabla V$ is negative, the single-field trajectory parametrized by $\phi$ yields a negative value for $c$. This situation can naturally arise in the interior of field space, as illustrated in Fig.~\ref{fig:fade}, where the scalar potential possesses a local maximum. If the scalar field lies to the left of this maximum, then $\nabla m$ and $\nabla V$ point in opposite directions.  

A model in which both dark matter (DM) and dark energy (DE) always fade simultaneously due to the dynamics of a single scalar field, with $c$ and $c'$ sharing the same sign, was studied in~\cite{Agrawal:2019dlm}. The setup we consider here is the same with the minor generalization of allowing for both same-sign and opposite-sign values of $c$ and $c'$.\footnote{Additionally, we do not assume the presence of a flat region in $m(\phi)$, which was assumed in~\cite{Agrawal:2019dlm} for technical reasons.}  In the case they have opposite signs matter(energy) still fade in the corresponding matter(energy) dominated erass, but increase in the energy (matter) dominated eras.   A positive (negative) value of $c$ corresponds to a scenario where the fading of dark energy accompanies a decreasing (increasing) tower scale, which in the dark dimension scenario reflects an expanding (contracting) fifth dimension.

Let us comment on how non-zero values of \( c \) and \( c' \) affect different epochs of cosmology. During the radiation-dominated era, the scalar field remains effectively frozen due to Hubble friction being set by the radiation energy density. As a result, the dynamics are largely unaffected in this epoch. However, during the matter-dominated era, the situation changes: as the scalar field rolls, the dark matter density decreases due to the exponential dependence of the mass \( m(\phi) \) on \( \phi \). Consequently, this epoch and any observations probing it (e.g.,\ DESI) will be most sensitive\footnote{Of course, DESI DR2 has data that overlaps in redshift with SN datasets during the dark energy-dominated era and hence has some sensitivity to dark energy-evolution also. However, the majority of data for DESI DR2 is in the dark matter-dominated era~\cite[Table II]{DESI:2025zgx}.} to the parameter \( c' \). On the other hand, the dark energy–dominated era is more sensitive to modifications of the scalar potential itself and therefore provides a probe of \( c \). In this epoch, the scalar field rolls in the direction of $\rm{sgn}(c)$ so that the dark energy density decreases due to the exponential dependence of the scalar potential \( V(\phi) \) on \( \phi \).  Therefore, the supernovae datasets which probe low redshifts should be expected to be most sensitive to \( c \). Thus, one can intuitively associate the dominant effects of \( c' \) and \( c \) with distinct cosmological epochs: the matter-dominated era and the dark energy–dominated era, respectively. For non-zero values of \( c \) and \( c' \), we expect the dark matter to fade in the matter-dominated era and dark energy to fade in the dark energy–dominated era, which is why we call this model the \textit{Fading Dark Sector} model. Note that if \( c \) and \( c' \) have opposite signs, the dark energy increases in the dark matter dominated era and the dark matter increases in the dark energy–dominated era which is a secondary effect in the expansion of the universe.

Components of the energy density driving the expansion of our universe are often modeled as perfect fluids. These are characterized by an equation of state parameter $w$ which is the ratio of the fluid pressure to its energy density. In general, the equation of state is time-dependent although it is constant for many common fluids like cold dark matter, non-relativistic baryons, massless radiation, and the cosmological constant.  The relation between energy density and expansion scale depends on the fluid type with the relation $\rho\propto a^{-3(1+w)}$. The cosmological constant, which has $w = -1$ and an energy density that does not dilute with the expansion of the universe, marks the limiting case of this behavior. No known fluids have $w < -1$ which would mean that the energy density increases with the expansion of the universe. This behavior violates the null energy condition and would require a more exotic fluid known as phantom.

In light of the above theoretical discussion, it is interesting that recent measurements made by the DESI collaboration (and analyzed in conjunction with CMB and supernova data and the recent results of DES) show evidence of an equation of state for dark energy that is less than $-1$ \cite{DESI:2025fii,Gialamas:2025pwv} and thus violates the null energy condition.

The key to understanding the phantom behavior is that the dark energy equation of state cannot be defined unambiguously in the presence of a DM/DE coupling. This is because, when such a coupling exists, the interaction term contributes to the energy momentum tensor. Moreover, the individual energy densities are no longer simple functions of the scale factor, and the naive definition $w = -\frac{1}{3}\frac{d\log(\rho_{\rm DE})}{d\log(a)}$ is influenced by the interaction. In particular, if one assumes that DM redshifts as $w = 0$ (i.e., like $1/a^3$), then one can subtract its contribution from the total energy density and fit the evolution of DE accordingly. In the presence of interactions between DM and DE, this effective equation of state is not constrained to satisfy $w > -1$, and in fact, for certain couplings, we can obtain $w < -1$. Since the new DESI data analysis assumes $\rho_{\rm DM} \sim 1/a^3$, we advocate that the evidence for $w_{\rm DE} < -1$ found by the DESI collaboration is actually evidence for a coupling between dark energy and dark matter.

It is easy to understand intuitively why a coupling between dark matter and dark energy can lead to an effective dark energy equation of state $w_{\mathrm{DE,eff}} < -1$, and we will see that this indeed occurs in our model. Consider the evolution of the combined DM + DE fluid in $\Lambda$CDM, where the total energy density redshifts in a prescribed manner. Now, suppose there is a coupling between DM and DE that causes this combined fluid to dilute more slowly than in $\Lambda$CDM. Heuristically, this can occur if DM transfers part of its energy density to DE, although, as noted earlier, it is not possible to define these energy densities separately and precisely in the presence of interactions. Nevertheless, if one models this interacting system under the assumption that DM still redshifts as $1/a^3$, then one is forced to infer the existence of a DE component with $w < -1$, since this is the only way to achieve slower dilution than in $\Lambda$CDM while keeping the matter dilution rate and density at early times fixed.

Let us illustrate the above behavior in our model. The total energy density and pressure of the combined dark matter and dark energy after the onset of the fading dark sector at $\phi=0$ are given by
\begin{align}
    \rho_{\rm DM+DE} &= m_0 n_0 a^{-3}e^{-c'\phi}+\frac{1}{2}\dot\phi^2+V_0e^{-c\phi}\,,\nonumber\\
    p_{\rm DM+DE} &= \frac{1}{2}\dot\phi^2 - V_0e^{-c\phi}\,,
\end{align}
where the constants $m_0$ and $n_0$ represent the initial mass scale and density of dark matter, while $V_0$ is a constant that sets the energy scale of the scalar potential. We set the beginning of the cosmological simulation, and hence the onset of the fading dark matter, to redshift $z = 10^{14}$. However, as shown in Fig.~\ref{fig:phi} the scalar field remains frozen by Hubble friction throughout most of the radiation-dominated era and only begins to roll significantly near its end. Therefore, the precise value of the initial redshift is not physically relevant.

\begin{figure*}[!tp]
    \centering
    \begin{subfigure}{.49\textwidth}
        \includegraphics[width=\linewidth]{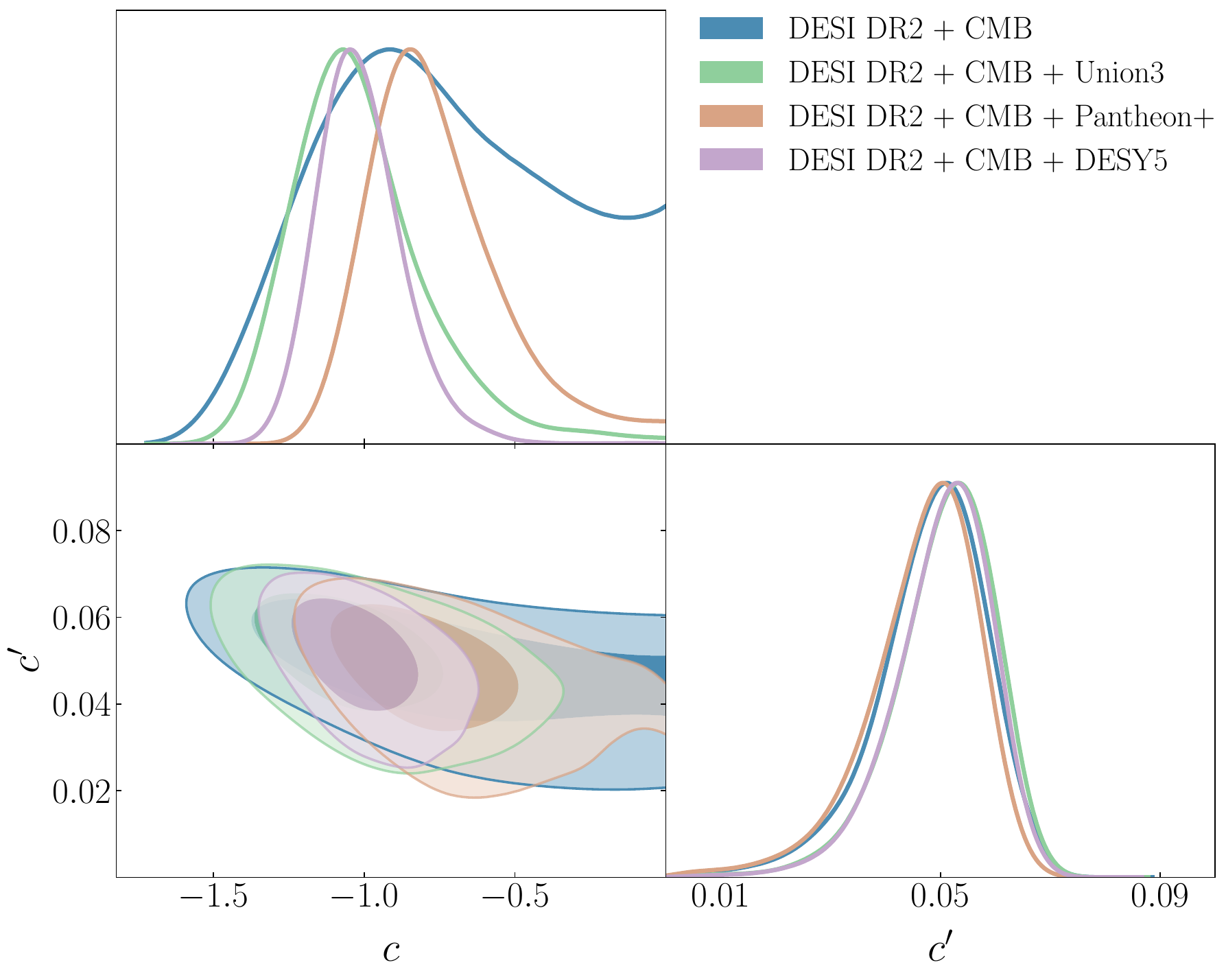}
        \caption{$c<0,c'>0$}
        \label{contour_cpcpn}
    \end{subfigure}
    \hfill
    \begin{subfigure}{.49\textwidth}
        \includegraphics[width=\linewidth]{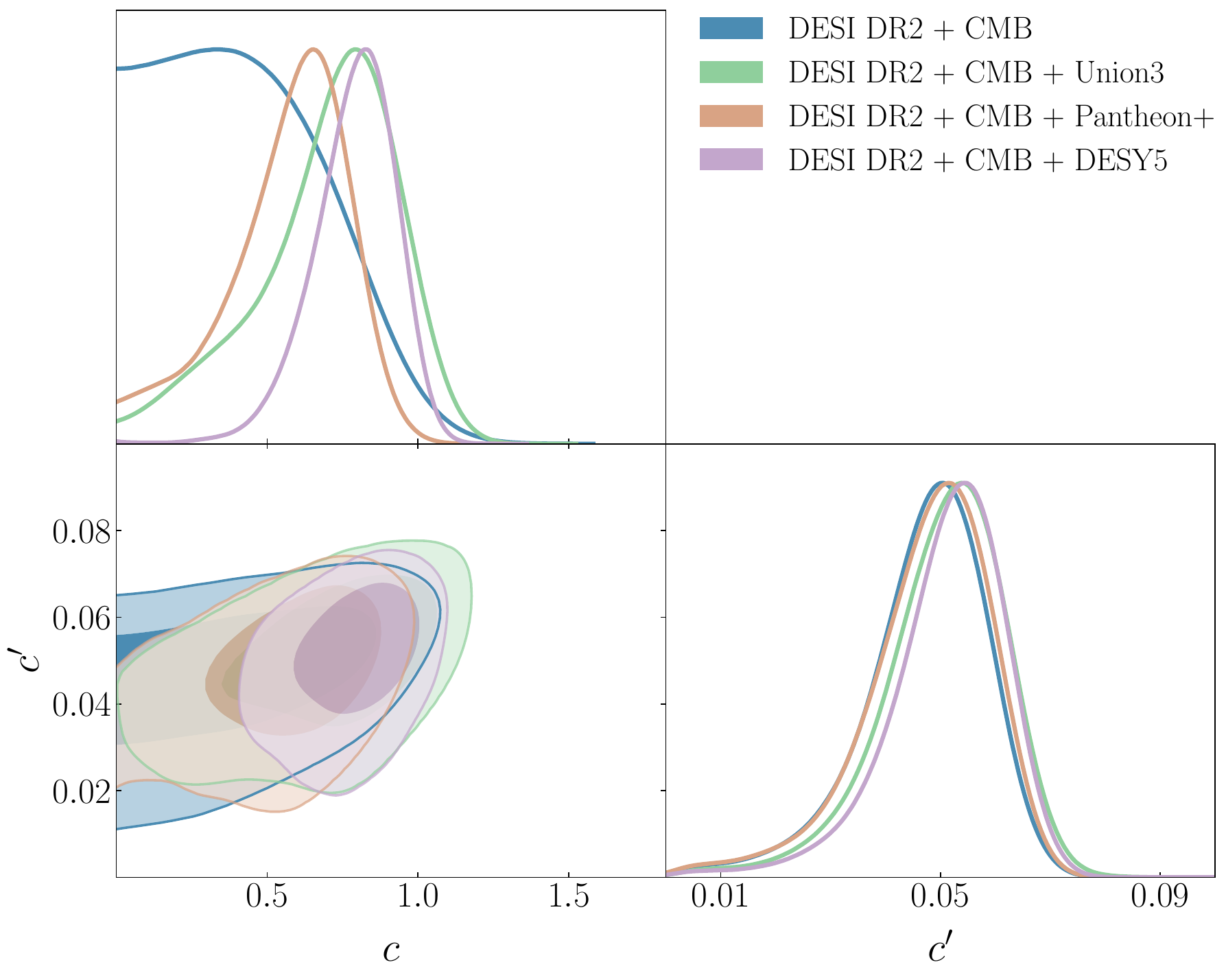}
        \caption{$c>0,c'>0$}
        \label{contour_cpcpp}
    \end{subfigure}
 \caption{Contours showing the 1- and 2-$\sigma$ constraints on $c$ and $c'$, along with their marginalized posteriors for two fading dark sector models with opposite signs of $c$, across various datasets.  $\Lambda$CDM corresponds to $c=c'=0$.}
    \label{Contours}
\end{figure*}

\begin{table*}[ht!]
\centering
\begin{tabular}{|c|>{\hspace{.2cm}}c<{\hspace{.2cm}}|>{\hspace{.2cm}}c<{\hspace{.2cm}}|c|c|>{\hspace{.2cm}}c<{\hspace{.2cm}}|c|c|}
\hline
\multirow{2}{*}{Datasets} & $w_0 w_a$CDM & \multicolumn{3}{c|}{FDS ($c<0$)}& \multicolumn{3}{c|}{FDS ($c>0$)}\\
\cline{2-8}
 & $\sigma$ & $\sigma$ & $c$ & $c'$ & $\sigma$ & $c$ & $c'$\\
\hline
\multirow{2}{*}{DESI+CMB} & \multirow{2}{*}{$3.0$} & \multirow{2}{*}{$2.5$} & $-1.03$ & $0.05$ & \multirow{2}{*}{$2.1$} & $0.42$& $0.05$\\
 &   &  & $-0.72^{+0.45}_{-0.43}$ & $0.05\pm0.01$ & & $0.43^{+0.28}_{-0.29}$ & $0.05\pm 0.01$\\ 
\arrayrulecolor{gray!30}\hline
\multirow{2}{*}{DESI+CMB+Union3} & \multirow{2}{*}{$3.7$} & \multirow{2}{*}{$3.4$} & $-1.12$ & $0.05$ & \multirow{2}{*}{$2.9$} & $0.81$ & $0.06$\\
 &  &  & $-1.01^{+0.21}_{-0.22}$ & $0.05\pm 0.01$ & & $0.71^{+0.22}_{-0.24}$ & $0.05\pm 0.01$\\ 
 \arrayrulecolor{gray!30}\hline
\multirow{2}{*}{DESI+CMB+Pantheon+} & \multirow{2}{*}{$2.7$} & \multirow{2}{*}{$2.9$} & $-0.85$ & $0.05$ & \multirow{2}{*}{$2.6$} & $0.65$ & $0.05$\\
 &  &  & $-0.76^{+0.23}_{-0.22}$ & $0.05\pm 0.01$ & & $0.54^{+0.21}_{-0.24}$ & $0.05\pm 0.01$\\ 
 \arrayrulecolor{gray!30}\hline
\multirow{2}{*}{DESI+CMB+DESY5} & \multirow{2}{*}{$4.1$} & \multirow{2}{*}{$4.0$} & $-1.06$ & $0.05$ & \multirow{2}{*}{$3.6$} & $0.83$ & $0.06$\\
 &  &  & $-1.00^{+0.13}_{-0.14}$ & $0.05\pm 0.01$ & & $0.79^{+0.14}_{-0.12}$ & $0.05\pm 0.01$\\ 
\arrayrulecolor{black}\hline
\end{tabular}
\caption{Statistical significance comparing the fading dark sector models for both $c > 0$ and $c < 0$ alongside the CPL parametrization, across various combinations of supernova datasets (including the recalibrated supernovae datasets), to $\Lambda$CDM. Additionally, the best fit, the mean, and the $\pm 1\sigma$ values for $c,c'$ are provided for each fading dark sector model.}
\label{sigmas}
\end{table*}

To extract the effective dark matter and dark energy components, we define the dark matter sector density and pressure as
\begin{align}
\label{eq:DM equations}
    \rho_{\rm DM} &= m_0 n_0e^{-c'\phi_i}a^{-3}=\rho^0_{\rm DM}e^{-c'\phi_i}a^{-3}\,,\nonumber\\
    p_{\rm DM} &= 0\,,
\end{align}
so that it redshifts exactly as conventional cold dark matter. The remainder is then attributed to the dark energy sector:
\begin{align}
    \rho_{\rm DE} &= \frac{1}{2}\dot\phi^2+V_0e^{-c\phi}+\frac{\rho^0_{\rm DM}\left[e^{-c'\phi}-e^{-c'\phi_i}\right]}{a^{3}}\,,\nonumber\\
    p_{\rm DE} &= \frac{1}{2}\dot\phi^2 - V_0e^{-c\phi}\,,
\end{align}
where $\phi_i$ is the value of the scalar field at the onset of the dark matter mass evolution in comoving volume.

Note that the dark matter density and pressure have been defined so that the dark matter behaves as a pressureless fluid redshifting as $1/a^3$, while all deviations due to the field dependence of $m$ are attributed to the dark energy sector. With the above definitions, it is easy to see that the effective DE equation of state can be less than one. To that end, and for convenience in the following discussion, we first define the scalar field equation of state:
\begin{align}
    w_\phi = \frac{p_\phi}{\rho_\phi} = \frac{\frac{1}{2}\dot\phi^2 - V_0e^{-c\phi}}{\frac{1}{2}\dot\phi^2 + V_0e^{-c\phi}}
\end{align}
which satisfies $-1 \leq w_\phi \leq 1$ with the limiting cases corresponding to a scalar field with vanishing kinetic energy and potential respectively. In term of the quantity $w_\phi$, we write the effective DE equation of state as:
\begin{align}\label{weff}
    w_{\rm {DE, eff}} = \frac{p_{\rm DE}}{\rho_{\rm DE}} = \frac{w_\phi}{1 + x} 
\end{align}
where
\begin{align}
    x \equiv \frac{\rho^0_{\rm DM}}{a^3 \rho_\phi} \left[e^{-c'\phi}-e^{-c'\phi_i}\right].
\end{align}
For a non-rolling scalar field, we have $\phi = \phi_i$ and $w_{\rm{DE, eff}} \rightarrow w_\phi$ satisfying the usual expectation that uncoupled quintessence DE has an equation of state between $-1$ and $1$. On the other hand, if the scalar field rolls to larger values as the universe expands, i.e. $\phi_i < \phi$, then we have that $x < 0$ and $w_{\rm {DE, eff}} < w_\phi$ which allows $w_{\rm {DE, eff}} < -1$ if $w_\phi$ is sufficiently close to $-1$. This is indeed the behaviour we get in the fading DM model. The equations of motion for the scale factor and the scalar field are
\begin{align}
    &3H^2=\frac{1}{2}\dot\phi^2+V_{\rm{eff}}(\phi)+\frac{\Omega_R^0}{a^4}(3H_0^2)+\frac{\Omega_{B}^0}{a^3}(3H_0^2)\,,\nonumber\\
    &\ddot\phi+3H\dot\phi+\frac{d V_{\rm{eff}}}{d\phi}=0\,,
\end{align}
 where $\Omega_{R}$ and $\Omega_{B}$ are the current ratios of the energy density of radiation and baryonic matter to the critical energy density and 
 \begin{align}
     V_{\rm{eff}}=V_0e^{-c\phi}+m_0n_0a^{-3}e^{-c'\phi}.
 \end{align}
The fact that $w_{\rm {DE, eff}}$ can have a magnification factor compared to $w_\phi$ and make it appear phantom-like was already noted in \cite{Stefancic:2003bj,Huey:2004qv,Das:2005yj,vandeBruck:2019vzd,Agrawal:2019dlm,Chakraborty:2025syu,Khoury:2025txd}.  All these works except for~\cite{Agrawal:2019dlm} assumed the mass of the dark matter increases as the scalar potential decreases, which was motivated in part by the condition that the effective potential acquires a minimum which is not required in~\cite{Agrawal:2019dlm}.

Another key difference is in the definition of the effective dark energy density. Except for~\cite{Agrawal:2019dlm}, to attain a phantom behavior despite the fact that mass goes up, they define an effective dark energy density by subtracting the redshifted contribution of the current dark matter density instead of the initial one\footnote{Since the sampling of the data is scattered across different epochs the effective observational $w_{\rm{DE,eff}}$ is presumably a subtraction of fixed dark matter density at some intermediate point.}. The combination of these two differences leads for those models to potentially also have an effective phantom behavior.  In our model, even with opposite $c,c'$ signs, the dark matter mass goes down in the matter dominated era and goes back up slightly in the energy dominated era, so the net effect for us is a decrease in the DM mass leading to a natural phantom behavior already discussed.  This is unlike e.g.,~\cite{Khoury:2025txd} where DM mass monotonically increases in DESI-sensitive era (though it can go down in earlier epochs).

Before proceeding further, we would like to take this opportunity to compare the theoretical structure of our model to those appearing in~\cite{Stefancic:2003bj,Huey:2004qv,Das:2005yj,vandeBruck:2019vzd,Chakraborty:2025syu,Khoury:2025txd}. We will use the most recent paper~\cite{Khoury:2025txd} as an example but our comments apply in general. While the model  we are studying here~\cite{Agrawal:2019dlm} has  phenomenological similarities with~\cite{Khoury:2025txd}, in that they couple dark matter and dark energy, there are several substantial and very important differences. First, as we discussed the coupling parameters $c$ and $c'$ are expected, by string-theoretic considerations, to take $\mathcal{O}(1)$ values and this is borne out in our inference from data.  For example apriori $0\leq  |\nabla V|\leq 1$  in Planck units.  But string theory motivated dS conjecture~\cite{Obied:2018sgi,Bedroya:2019snp} led to the prediction~\cite{Agrawal:2018own,Agrawal:2019dlm} predicted $|\nabla V|\sim 10^{-122\pm 1}$ where the $\pm 1$ in the exponent is determined by $c$.  No analogous theoretical motivation is provided in~\cite{Khoury:2025txd}. Secondly for dark matter mass, string theoretic motivations led \cite{Agrawal:2019dlm}
to propose that the slope of dark matter mass $|\nabla_\phi m|$, is of the order of the dark matter mass $m$  (which again a priori could have been any number from 0 to 1 in Planck units).  By contrast, in~\cite{Khoury:2025txd} only a correction to the dark matter mass is $\phi$-dependent. Achieving an observable effect in the latter case requires the correction term to be comparable to the bare mass, introducing an additional tuning. Finally, our model is characterized by only two additional $O(1)$ parameters, $c$ and $c'$, while the model in~\cite{Khoury:2025txd} contains three additional free parameters. As a result, despite certain phenomenological similarities, the theoretical structure underlying the two models is substantially different.

\begin{figure*}[!tp]
    \centering
    \begin{subfigure}{\textwidth}
        \includegraphics[width=.49\linewidth]{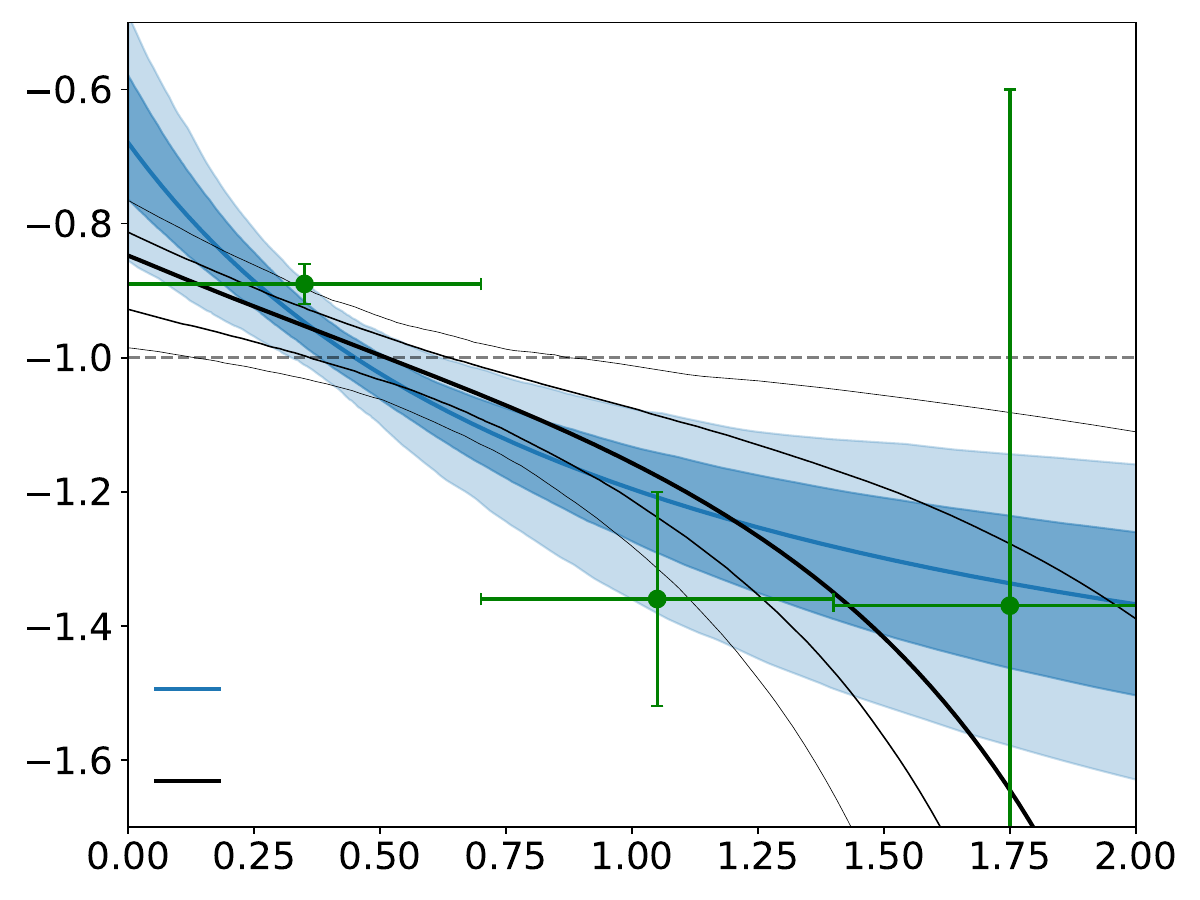}
        \hfill
        \includegraphics[width=.49\linewidth]{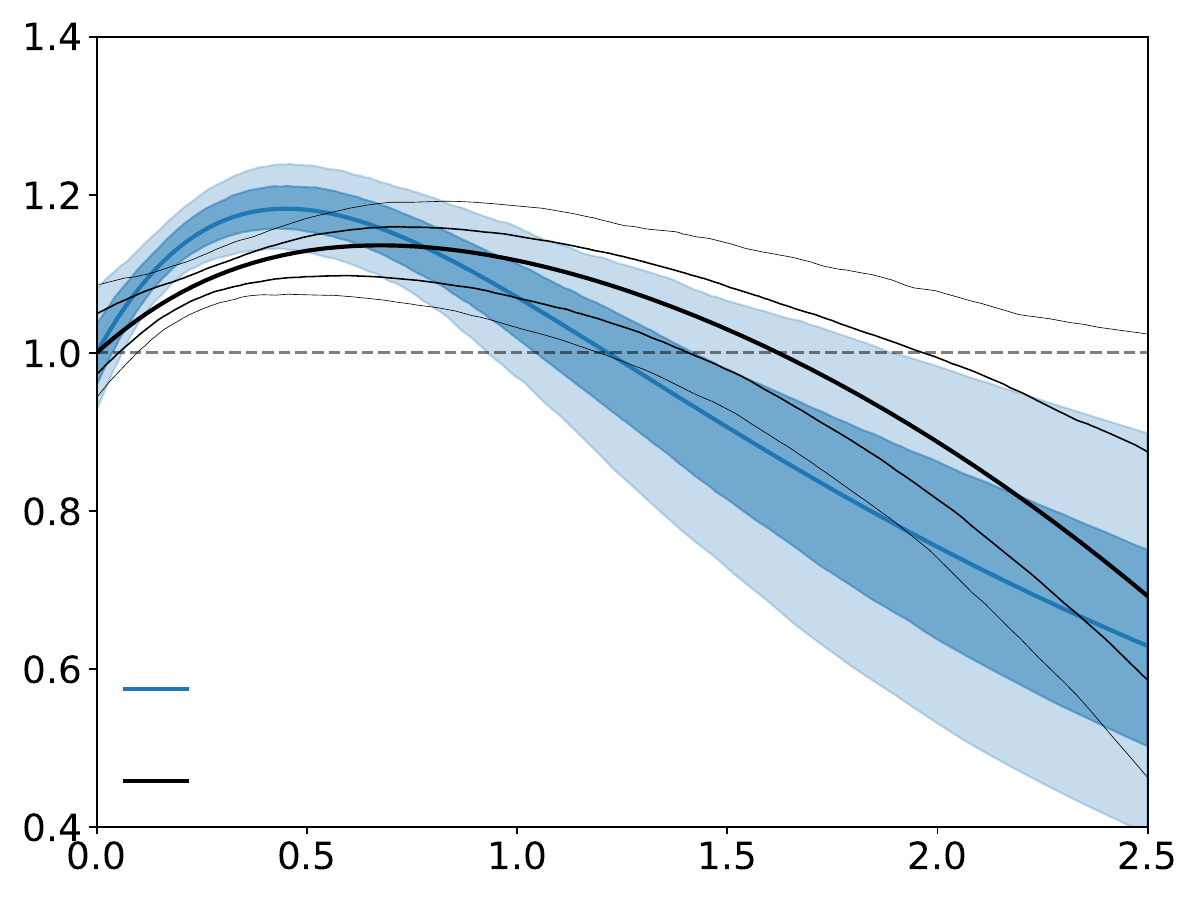}
        \begin{picture}(0,0)\vspace*{-1.2cm}
            \put(-515,80){\rotatebox{90}{$w_{\rm DE,eff}$}}
            \put(-380,-5){$z$}
            \put(-265,80){\rotatebox{90}{$f_{\rm DE,eff}(z)$}}
            \put(-125,-5){$z$}
            \put(-210,41){CPL}
            \put(-210,23){FDS}
            \put(-460,41){CPL}
            \put(-460,23){FDS}
        \end{picture}\vspace*{0.25cm}

        \caption{$c<0,c'>0$}
    \end{subfigure}
    \hfill
    \begin{subfigure}{\textwidth}
        \includegraphics[width=.49\linewidth]{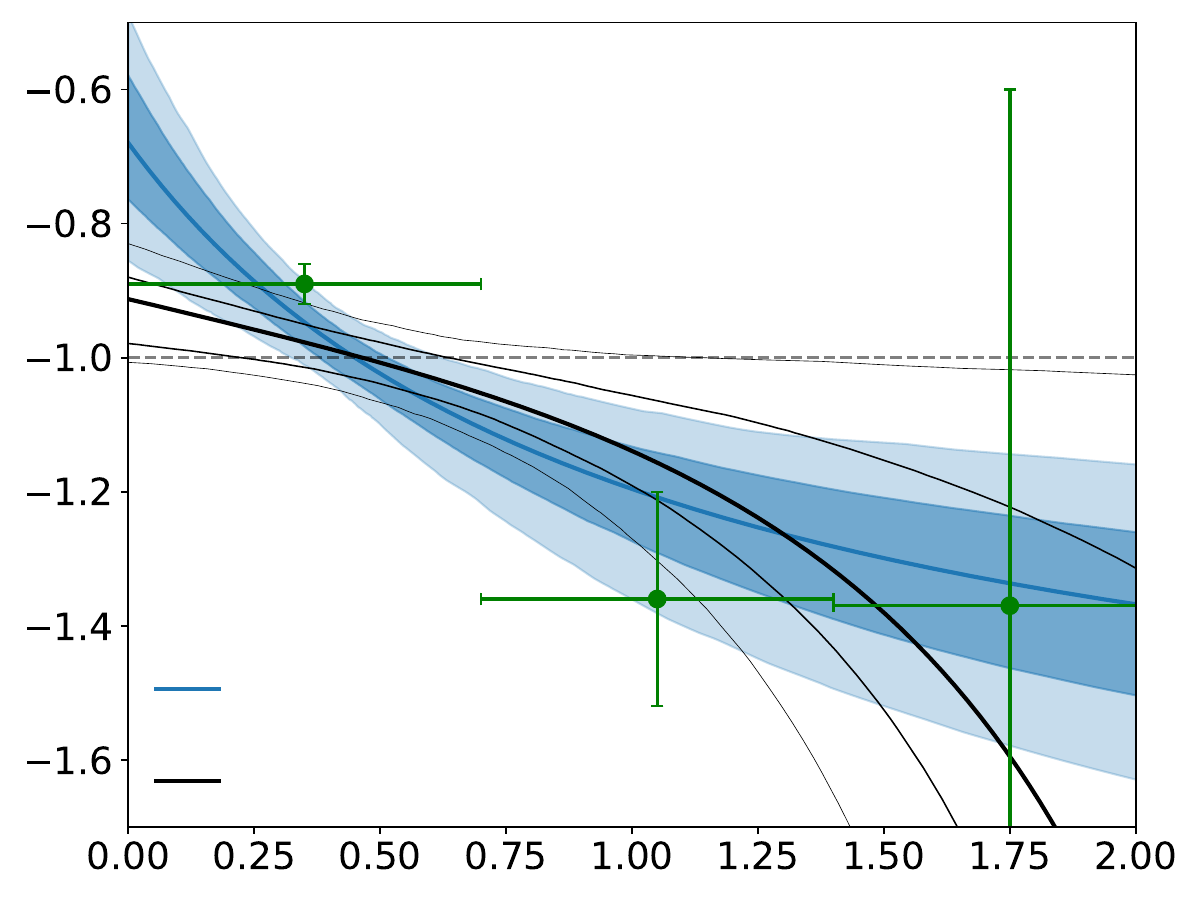}
        \hfill
        \includegraphics[width=.49\linewidth]{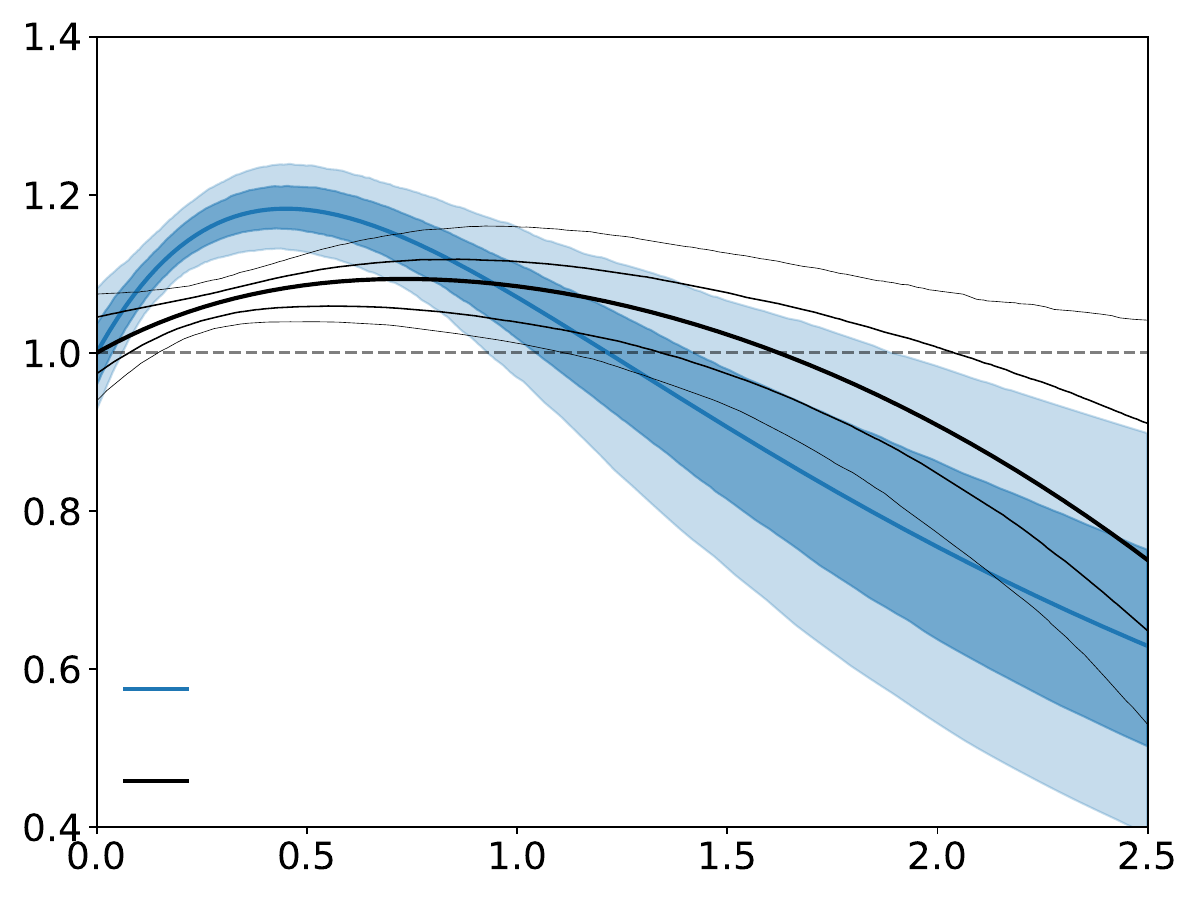}
        \begin{picture}(0,0)\vspace*{-1.2cm}
            \put(-515,80){\rotatebox{90}{$w_{\rm DE,eff}$}}
            \put(-380,-5){$z$}
            \put(-265,80){\rotatebox{90}{$f_{\rm DE,eff}(z)$}}
            \put(-125,-5){$z$}
            \put(-210,41){CPL}
            \put(-210,23){FDS}
            \put(-460,41){CPL}
            \put(-460,23){FDS}
        \end{picture}\vspace*{0.25cm}
        \caption{$c>0,c'>0$}
    \end{subfigure}
    \caption{Comparison between (effective) dark energy of the fading dark sector models and the CPL parameterization using MCMC runs fitted to CMB + DESI DR2 + Union3 to remain comparable to that demonstrated in~\cite{DESI:2025fii}. In the left figures, we show the best fit, the 1- and 2-$\sigma$ effective dark energy equation of state $w_{\rm DE, eff}$ for two fading dark sector models with different signs of $c$ and compare both to the CPL parameterization plotted against the binned $w(z)$ from DESI DR2 (reproduced from~\cite[Fig.~12]{DESI:2025zgx} and shown in green). As illustrated, apparent phantom behavior can emerge in the fading dark sector scenario. In the right figures, we have also shown the best fit, the 1- and 2-$\sigma$ ranges for the normalized dark energy density, defined as $f_{\rm DE,eff}(z)=\rho_{\rm DE,eff}(z)/\rho_{\rm DE,eff}^{\rm best-fit}(0)$ for the two fading dark sector models.}
\label{wplot}
\end{figure*}

Given that our model is based on the fading dark matter model of~\cite{Agrawal:2019dlm} and that earlier results there show its consistency with the best-fit $w_0$–$w_a$ parameterization, we expect that it can reproduce the phantom-like behavior preferred by the DESI+CMB(+SN) datasets within a well-motivated physical framework. To explore this, we re-analyze the fading dark matter model using the newly available data. In order to run Markov Chain Monte Carlo (MCMC) simulations, we adapt the publicly available Boltzmann solver~\texttt{CLASS}~\cite{2011JCAP...07..034B} to include scalar fields coupled to a single species of Dark Matter (DM). We modify the evolution of the background cosmology and also include the new (coupled) perturbations in a self-consistent manner so that our modeling fully takes into account the effect of perturbations on linear scales. The equations governing the perturbations of this coupled system are standard in the literature, see e.g.,~\cite{Amendola:1999dr,Miranda:2017rdk}. The couplings, potentials, and dependence of the DM mass on the scalar fields (in the background and perturbation modules) are implemented in a way that is easy to modify and we use this freedom to test various models (such as those discussed in the Supplemental Materials~\ref{sec:supplemental}). 

We use the Metropolis--Hastings algorithm built into the sampler \texttt{Cobaya}~\cite{2019ascl.soft10019T,Torrado:2020dgo} to sample from the parameter posteriors in each case. Our MCMCs are run including the following datasets:
\begin{itemize}
    \item Baryon Acoustic Oscillations (BAO) measurements from DESI DR2~\cite{DESI:2025zgx}, which provide measurements of the transverse comoving distance $D_M(z) = \int_0^z \mathrm{d}z'\, H(z')^{-1}$ and the line-of-sight Hubble distance $D_H(z) = H(z)^{-1}$ for various tracers. 
    \item CMB temperature and polarization data from Planck 2018~\cite{Planck:2018vyg} alongside additional data from the Atacama Cosmology Telescope (ACT)~\cite{ACT:2025fju} for lensing and also from the latest \texttt{NPIPE} PR4 data release from the Planck collaboration~\cite{Carron:2022eyg}. These are made available through the likelihoods and baseline power spectra in the third Planck public data release (PR3), through the sixth data release (DR6) of the ACT collaboration, and through the \texttt{CamSpec} likelihoods~\cite{Rosenberg:2022sdy} integrated into the public version of \texttt{Cobaya}.\footnote{The ACT DR6 lensing likelihoods can be accessed through \href{https://github.com/ACTCollaboration/act_dr6_lenslike}{https://github.com/ACTCollaboration/act\_dr6\_lenslike}}
    \item Type Ia supernova luminosity distances. Here we use three supernovae samples: (i) the Union3 sample~\cite{Rubin:2023ovl}, (ii) the Pantheon+ sample~\cite{Scolnic:2021amr}, and (iii) the DESY5 sample~\cite{DES:2024jxu}. 
\end{itemize}

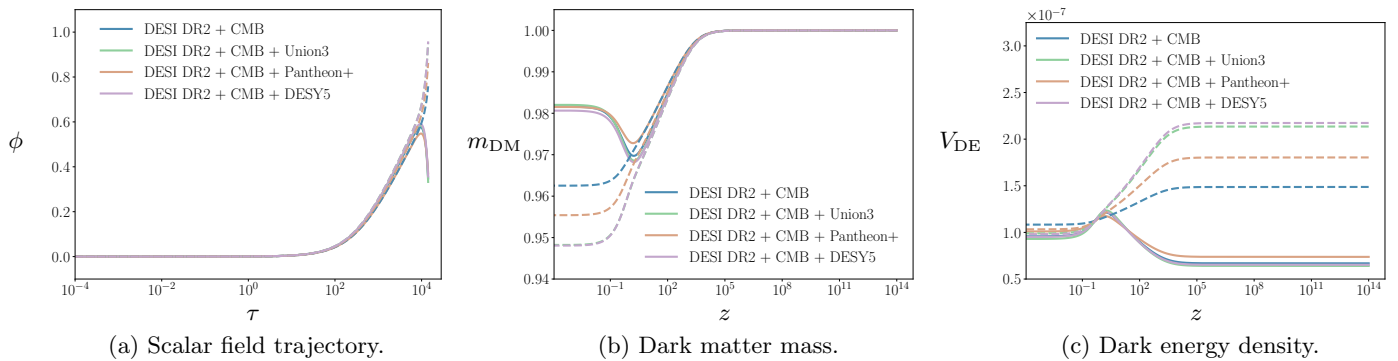
\begin{figure*}
    \centering
    
\begin{subfigure}{.3\textwidth}
\centering
    \includegraphics[width=\linewidth]{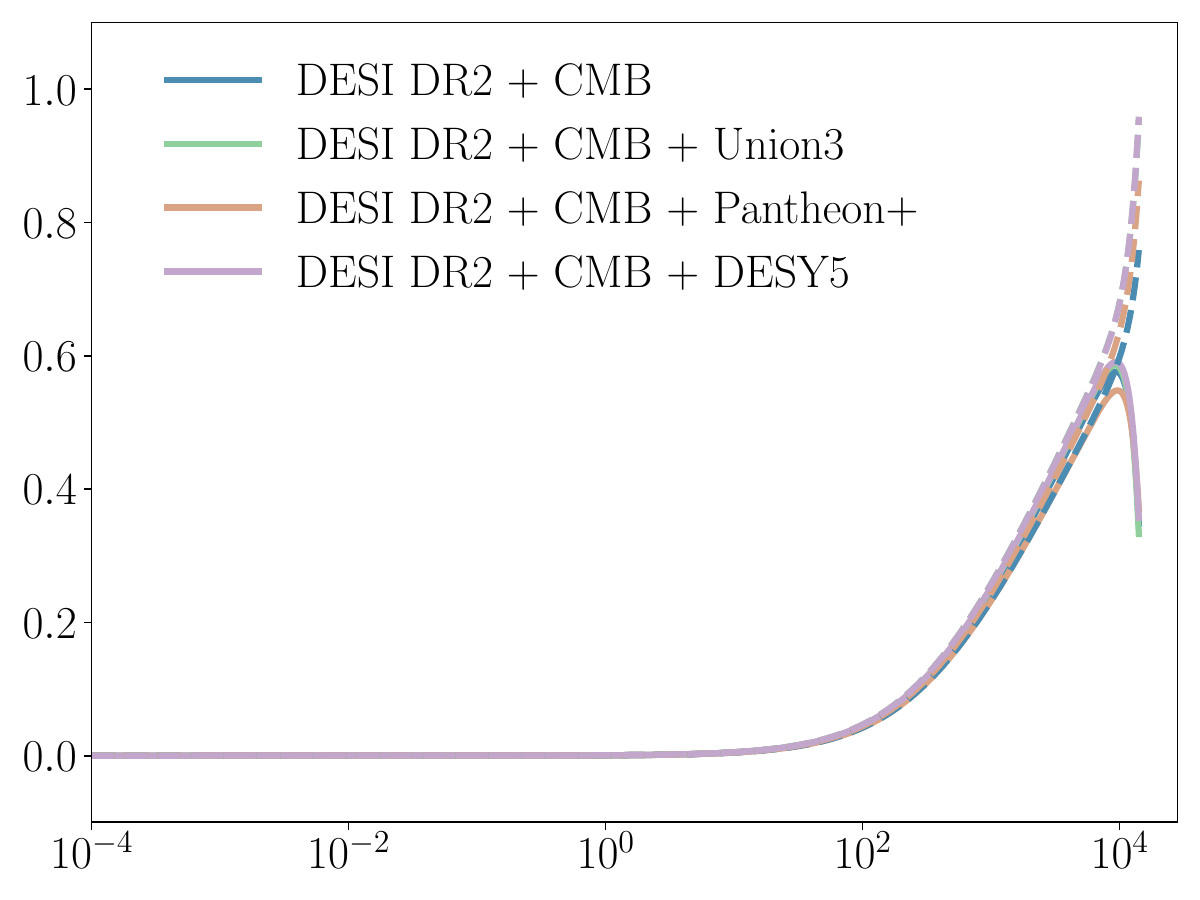}
    \begin{picture}(0,0)\vspace*{-1.2cm}
        \put(-90,70){{$\phi$}}
        \put(0,5){$\tau$}
    \end{picture}\vspace*{-0.2cm}
    \caption{Scalar field trajectory.}
\end{subfigure}
\hfill
\begin{subfigure}{.3\textwidth}
\centering
    \includegraphics[width=\linewidth]{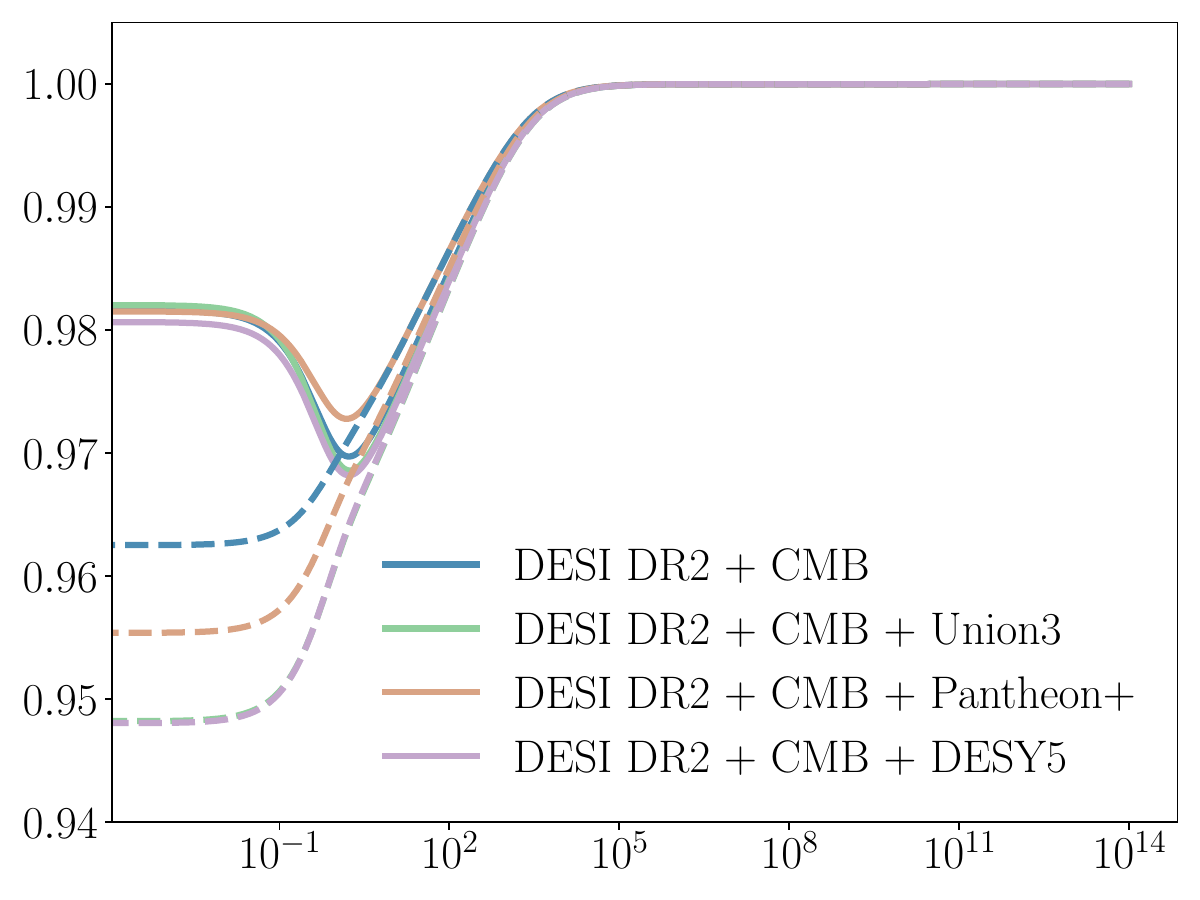}
    \begin{picture}(0,0)\vspace*{-1.2cm}
        \put(-95,70){{$m_{\rm DM}$}}
        \put(0,5){$z$}
    \end{picture}\vspace*{-0.2cm}
    \caption{Dark matter mass.}
\end{subfigure}
\hfill
\begin{subfigure}{.3\textwidth}
    \includegraphics[width=\linewidth]{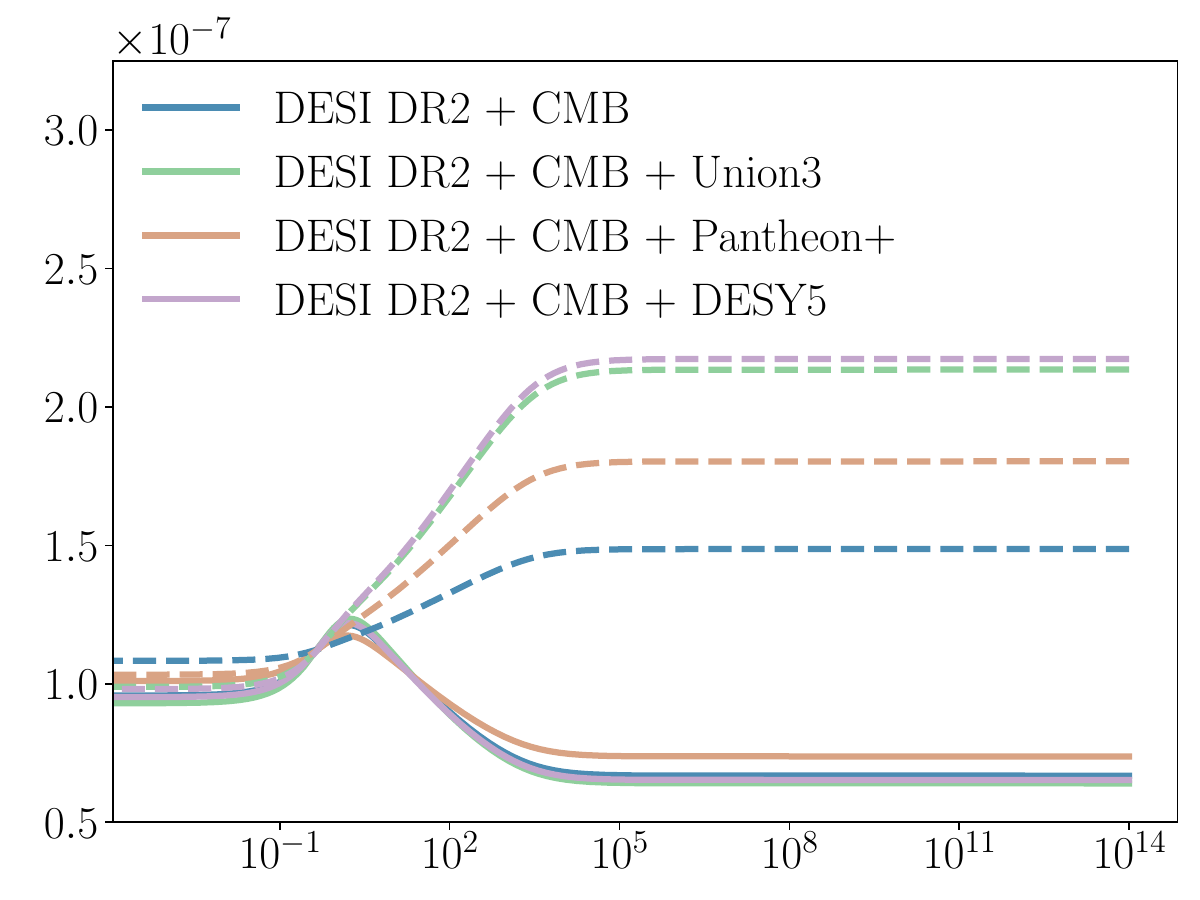}
    \begin{picture}(0,0)\vspace*{-1.2cm}
        \put(-95,70){{$V_{\rm DE}$}}
        \put(0,5){$z$}
    \end{picture}\vspace*{-0.2cm}
    \caption{Dark energy density.}
\end{subfigure}
\caption{Evolution of the scalar field and normalized mass scale of the tower that makes up the dark matter for two fading dark sector models with opposite signs of $c$, shown across various datasets. Here, the solid and dashed lines indicate the $c<0$ and $c>0$ FDS model, respectively. In both cases, the dark matter mass exhibits a net decrease. However, for $c<0$, there is a brief increase during dark energy domination due to the reversal in direction of the scalar field that controls the dark matter mass. We also show the potential of the dark energy for both of the FDS models.}

    \label{fig:phi}
\end{figure*}

\begin{figure*}[htp!]
    \centering
    \begin{subfigure}{\textwidth}
        \includegraphics[width=.49\textwidth]{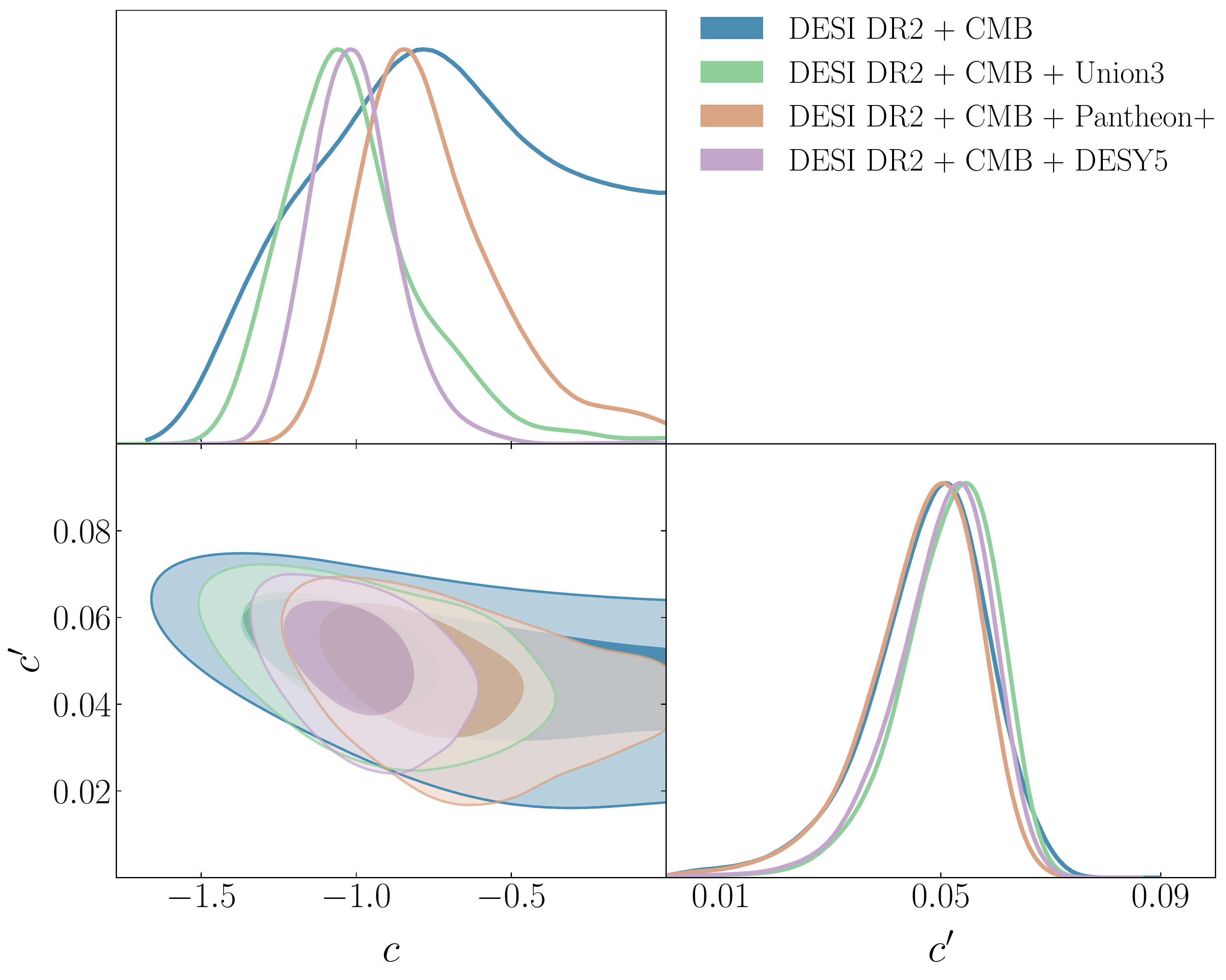}
        \hfill 
        \includegraphics[width=.49\textwidth]{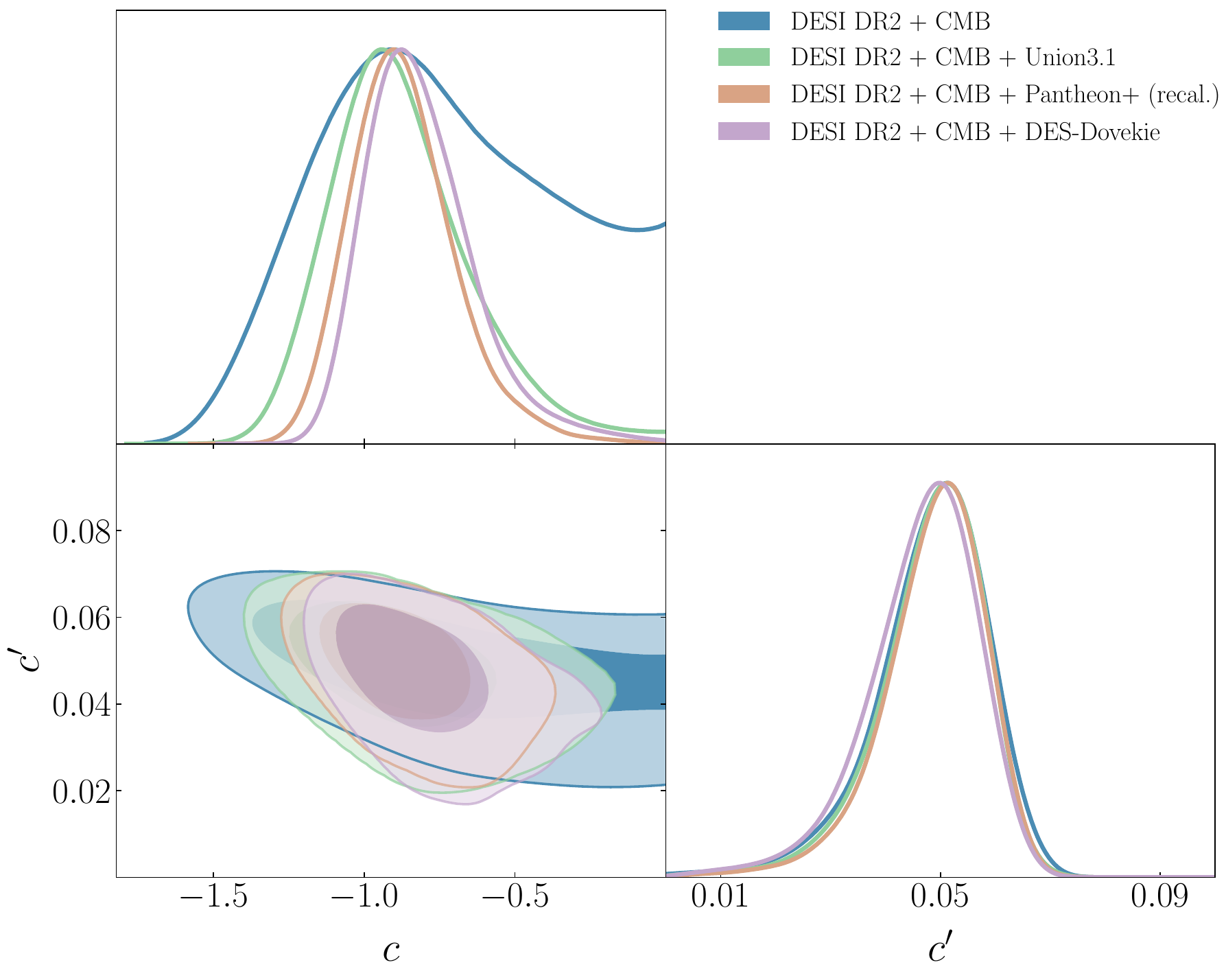}
        \begin{picture}(0,0)\vspace*{-1.2cm}
            \put(-300,140){\textcolor{red}{recalibrated}}
            \put(-300,130){\textcolor{red}{\vector(1,0){50}}}
        \end{picture}
        \caption{$c<0,c'>0$}
        \label{contour_cpcpn}
    \end{subfigure}
    \hfill\vspace{0.5cm}
    \begin{subfigure}{\textwidth}
        \includegraphics[width=.49\textwidth]{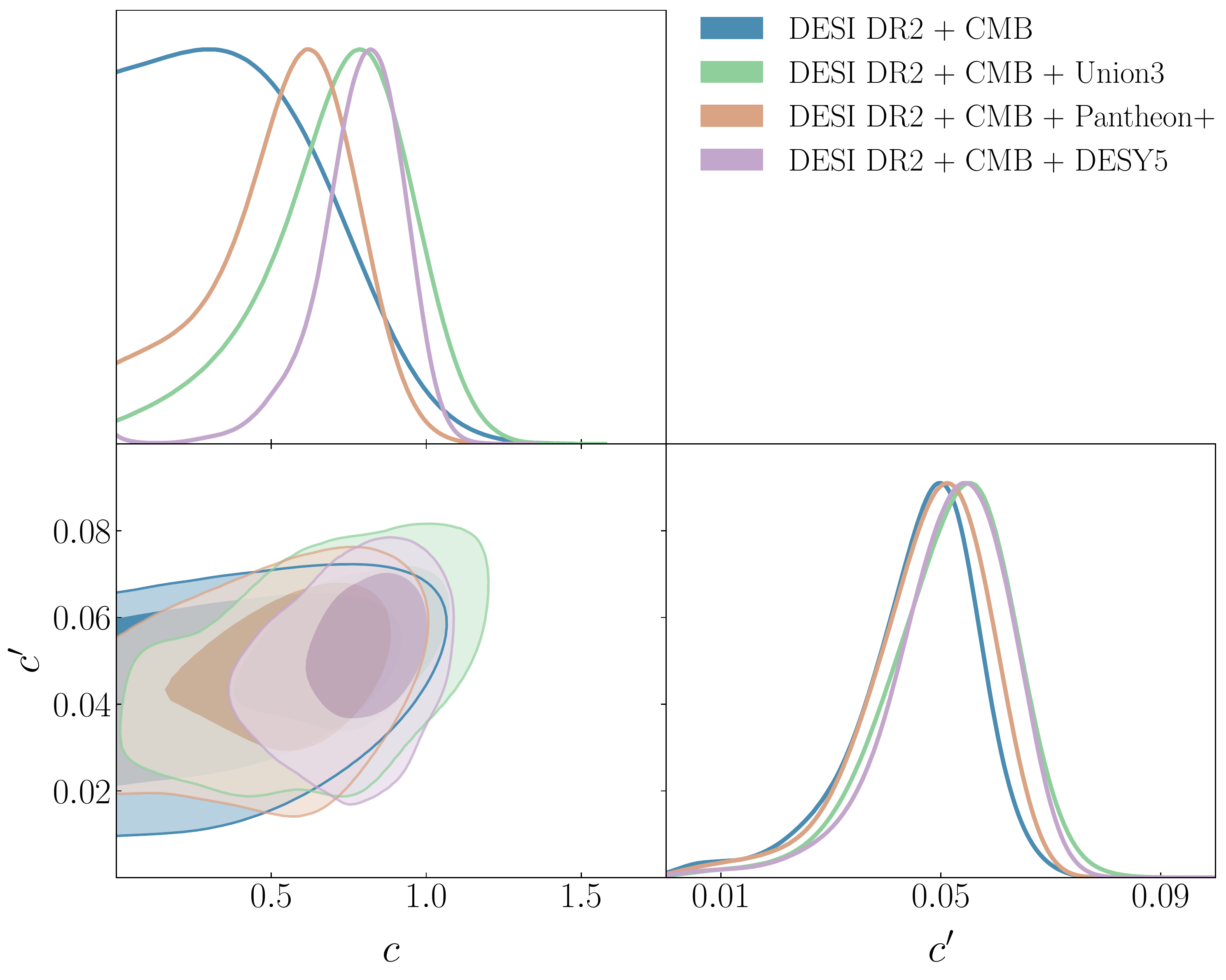}
        \hfill
        \includegraphics[width=.49\textwidth]{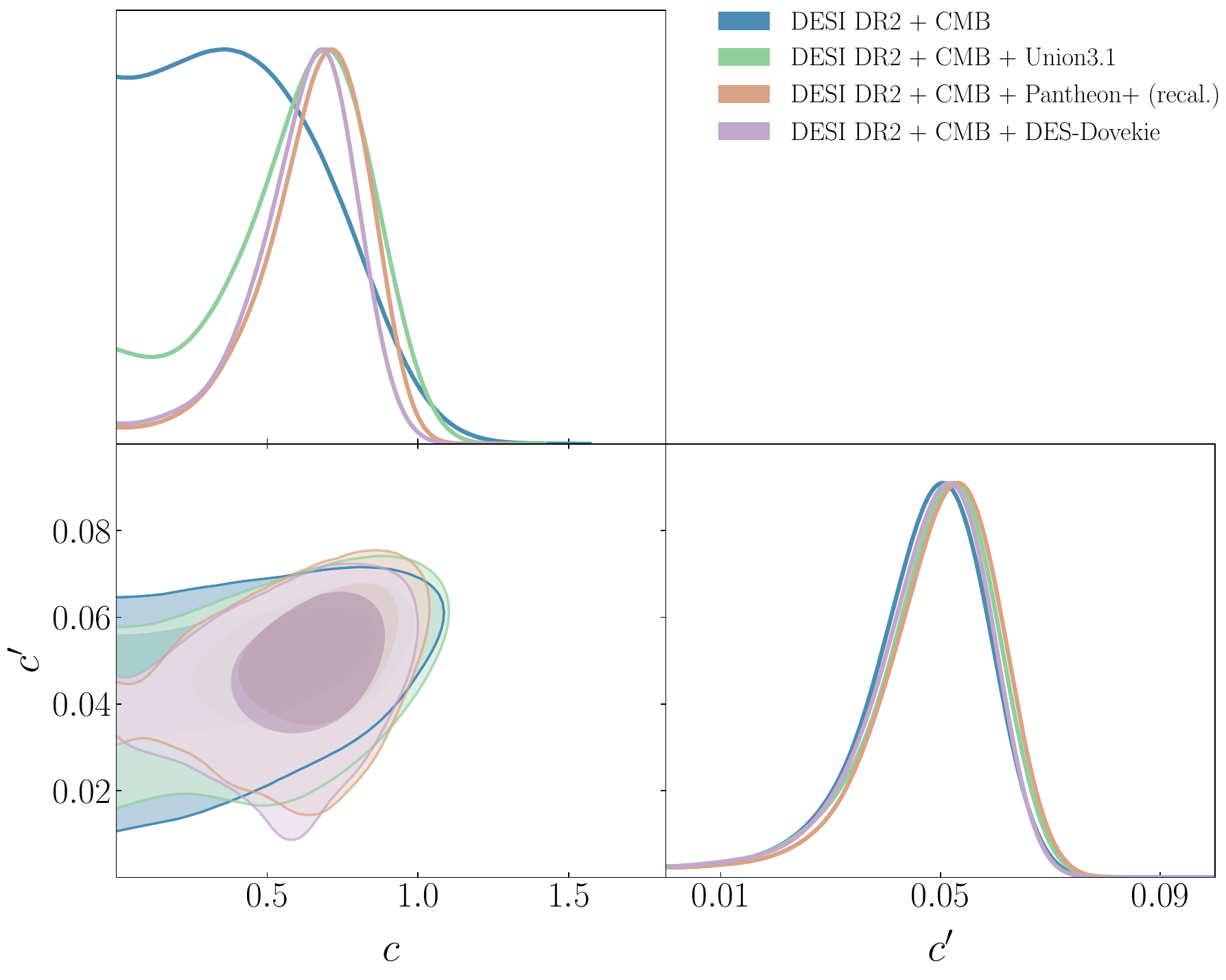}
        \begin{picture}(0,0)\vspace*{-1.2cm}
            \put(-300,140){\textcolor{red}{recalibrated}}
            \put(-300,130){\textcolor{red}{\vector(1,0){50}}}
        \end{picture}
        \caption{$c>0,c'>0$}
        \label{contour_cpcpp}
    \end{subfigure}
    \vspace{-0.5cm}
 \caption{Updated contours showing the 1- and 2-$\sigma$ constraints on $c$ and $c'$, along with their marginalized posteriors for two fading dark sector models with opposite signs of $c$, using the recalibrated supernovae datasets  $\Lambda$CDM corresponds to $c=c'=0$.}
    \label{Contours new}
\end{figure*}

\begin{figure*}[htp!]
    \centering
    \includegraphics[width=.49\linewidth]{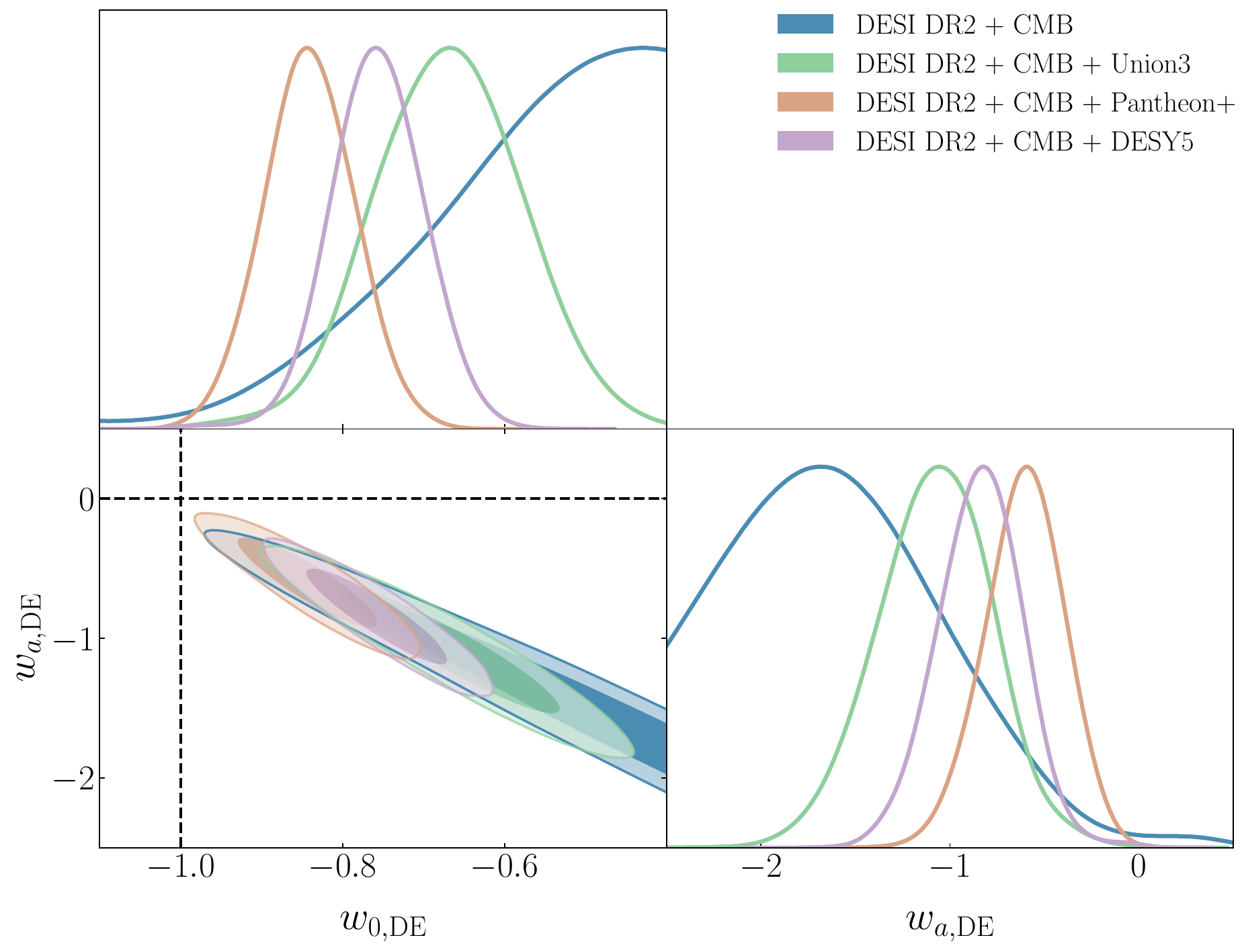}
    \hfill
    \includegraphics[width=.49\linewidth]{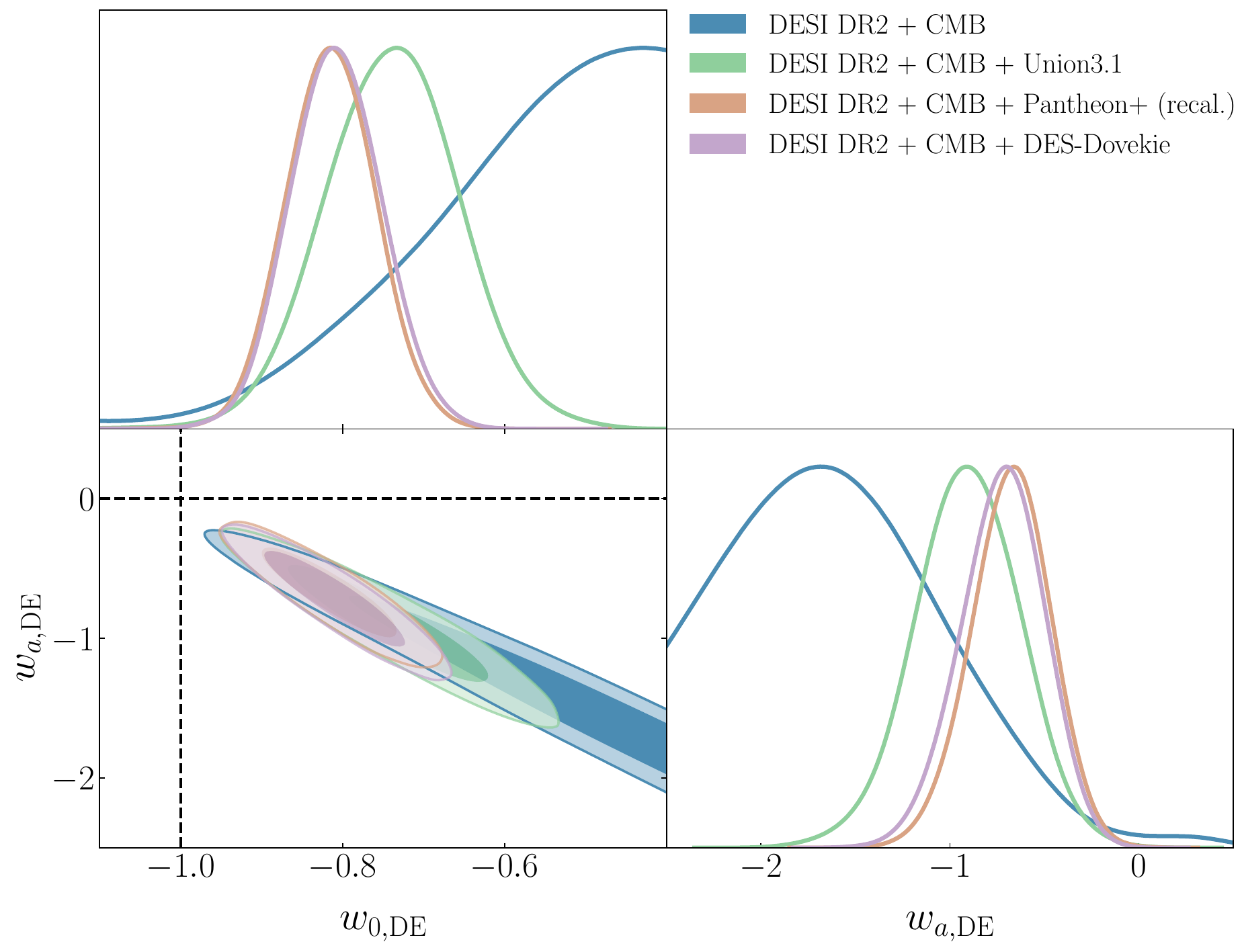}
    \begin{picture}(0,0)\vspace*{-1.2cm}
        \put(-300,140){\textcolor{red}{recalibrated}}
        \put(-300,130){\textcolor{red}{\vector(1,0){50}}}
    \end{picture}
    \caption{The updated contours showing the 1- and 2-$\sigma$ constraints on $w_0$ and $w_a$ in the CPL parameterization fitted to combinations of recently recalibrated supernovae datasets.}
    \label{fig:w0wa contour new}
\end{figure*}

We consider two models, distinguished by the sign of $c$, and analyze various combinations of cosmological datasets. These are two-parameter models characterized by either $c \geq 0,\ c' \geq 0$ or $c \leq 0,\ c' \geq 0$. The posterior contours and preferred values for $c$ and $c'$ across different datasets are shown in Fig.~\ref{Contours}. In both scenarios, during the dark matter–dominated era, the scalar field evolution is driven by the $\phi$-dependence of the tower, which pushes the field towards larger $\phi$, resulting in a decreasing dark matter mass. This behavior is illustrated in Fig.~\ref{fig:phi}, where we plot the evolution of the scalar field and dark matter density versus conformal time and redshift. The key difference between the two models arises during the transition to dark energy domination: for $c > 0$, the dark matter continues to decrease but with a modified rate, whereas for $c < 0$, the scalar field reverses direction and dark matter transitions from decreasing to increasing at late times. As shown in Table~\ref{sigmas}, our models have similar significance of preference over $\Lambda$CDM as the (unphysical) CPL parametrization. Between our two models, there is a mild statistical preference for the $c < 0$ model. However, regardless of the sign of $c$, all combination of datasets that include CMB+DESI yield consistent values for $c'$ and exhibit a strong preference for a non-zero $c' \simeq 0.05\pm 0.01$ (see Fig.~\ref{Contours}). Another remarkable feature is that any combination of datasets that include CMB+supernova consistently yield $\mathcal{O}(1)$ values for $|c|$ close to the Swampland bound $|c| \geq \sqrt{2/3}\approx 0.82$~\eqref{intTCC} (especially for $c<0$ case) which is expected to hold in some average sense in the interior of the field space (see Table~\ref{sigmas}). The contour plots for other cosmological parameters can be found in Fig.~\ref{fig:cpcpn} and Fig.~\ref{fig:cpcpp}.

\begin{table*}[ht!]
\centering
\begin{tabular}{|c|>{\hspace{.2cm}}c<{\hspace{.2cm}}|>{\hspace{.2cm}}c<{\hspace{.2cm}}|c|c|>{\hspace{.2cm}}c<{\hspace{.2cm}}|c|c|}
\hline
\multirow{2}{*}{Datasets} & $w_0 w_a$CDM & \multicolumn{3}{c|}{FDS ($c<0$)}& \multicolumn{3}{c|}{FDS ($c>0$)}\\
\cline{2-8}
 & $\sigma$ & $\sigma$ & $c$ & $c'$ & $\sigma$ & $c$ & $c'$\\
\hline
\multirow{2}{*}{DESI+CMB} & \multirow{2}{*}{$3.0$} & \multirow{2}{*}{$2.5$} & $-1.03$ & $0.05$ & \multirow{2}{*}{$2.1$} & $0.42$& $0.05$\\
 &   &  & $-0.72^{+0.45}_{-0.43}$ & $0.05\pm0.01$ & & $0.43^{+0.28}_{-0.29}$ & $0.05\pm 0.01$\\ 
\arrayrulecolor{gray!100}\hline
\multirow{2}{*}{DESI+CMB+Union3} & \multirow{2}{*}{$3.7$} & \multirow{2}{*}{$3.4$} & $-1.12$ & $0.05$ & \multirow{2}{*}{$2.9$} & $0.81$ & $0.06$\\
 &  &  & $-1.01^{+0.21}_{-0.22}$ & $0.05\pm 0.01$ & & $0.71^{+0.22}_{-0.24}$ & $0.05\pm 0.01$\\ 
 \arrayrulecolor{gray!30}\hline
 \multirow{2}{*}{ DESI+CMB+Union3.1} & \multirow{2}{*}{$3.4$} & \multirow{2}{*}{$3.0$} & $-0.98$ & $0.05$ & \multirow{2}{*}{$2.5$} & $0.72$ & $0.05$\\
  &  &  & $-0.87\pm 0.24$ & $0.05\pm 0.01$ & & $0.60\pm 0.24$ & $0.05\pm 0.01$\\ 
 \arrayrulecolor{gray!100}\hline
\multirow{2}{*}{DESI+CMB+Pantheon+} & \multirow{2}{*}{$2.7$} & \multirow{2}{*}{$2.9$} & $-0.85$ & $0.05$ & \multirow{2}{*}{$2.6$} & $0.65$ & $0.05$\\
 &  &  & $-0.76^{+0.23}_{-0.22}$ & $0.05\pm 0.01$ & & $0.54^{+0.21}_{-0.24}$ & $0.05\pm 0.01$\\ 
 \arrayrulecolor{gray!30}\hline
 \multirow{2}{*}{DESI+CMB+Pantheon+(recal.)} & \multirow{2}{*}{$3.2$} & \multirow{2}{*}{$3.2$} & $-0.92$ & $0.05$ & \multirow{2}{*}{$2.9$} & $0.73$ & $0.05$\\
 &  &  & $-0.87\pm 0.18$ & $0.05\pm 0.01$ & & $0.65\pm 0.18$ & $0.05\pm 0.01$\\ 
 \arrayrulecolor{gray!100}\hline
\multirow{2}{*}{DESI+CMB+DESY5} & \multirow{2}{*}{$4.1$} & \multirow{2}{*}{$4.0$} & $-1.06$ & $0.05$ & \multirow{2}{*}{$3.6$} & $0.83$ & $0.06$\\
 &  &  & $-1.00^{+0.13}_{-0.14}$ & $0.05\pm 0.01$ & & $0.79^{+0.14}_{-0.12}$ & $0.05\pm 0.01$\\ 
 \arrayrulecolor{gray!30}\hline
 \multirow{2}{*}{DESI+CMB+DES-Dovekie} & \multirow{2}{*}{$3.1$} & \multirow{2}{*}{$3.2$} & $-0.92$ & $0.05$ & \multirow{2}{*}{$2.8$} & $0.69$ & $0.05$\\
 &  &  & $-0.81\pm 0.19$ & $0.05\pm 0.01$ & & $0.62\pm 0.18$ & $0.05\pm 0.01$\\ 
\arrayrulecolor{black}\hline
\end{tabular}
\begin{picture}(0,0)\vspace*{-1.2cm}
    \put(-290,30){\textcolor{red}{\vector(0,-1){10}}}
    \put(-290,-13){\textcolor{red}{\vector(0,-1){10}}}
    \put(-290,-57){\textcolor{red}{\vector(0,-1){10}}}
    \put(-246,30){\textcolor{red}{\vector(0,-1){10}}}
    \put(-246,-13){\textcolor{red}{\vector(0,-1){10}}}
    \put(-246,-57){\textcolor{red}{\vector(0,-1){10}}}
    \put(-115,30){\textcolor{red}{\vector(0,-1){10}}}
    \put(-115,-13){\textcolor{red}{\vector(0,-1){10}}}
    \put(-115,-57){\textcolor{red}{\vector(0,-1){10}}}
\end{picture}
\caption{The updated statistical significance comparing the fading dark sector models for both $c > 0$ and $c < 0$ alongside the CPL parametrization, across various combinations of supernova datasets including the recalibrated supernovae datasets, to $\Lambda$CDM. As before, the best fit, the mean, and the $\pm 1\sigma$ values for $c,c'$ are provided for each fading dark sector model. The red arrow indicates how the significance has changed as the supernovae data gets recalibrated.}
\label{sigmas new}
\end{table*}

As shown in Fig.~\ref{wplot}, our models can also account for the apparent phantom behavior of dark energy reported by DESI DR2~\cite{DESI:2025fii}. We emphasize that $w$ is a derived, model-dependent quantity rather than a fundamental one, especially in scenarios involving couplings between dark matter and dark energy. For instance, although our model fits the data remarkably well, the effective equation of state $w_{\rm eff}$ diverges near redshift $z \simeq 5$ because the denominator in its definition (Eq.~\eqref{weff}) vanishes.
Additionally, similar to the CPL parameterization, the normalized dark energy density in the FDS model exhibits similar behavior that can capture qualitatively similar behaviors arising from the phantom crossing of $w_{\rm DE,eff}$ as can be seen in Fig.~\ref{wplot}. This is a general phenomena and persists even when fitting models to datasets involving recently recalibrated supernovae data as can be seen in Fig.~\ref{fig:updated weff}.

As we explained in Section~\ref{sec:4}, experiments probing the dark matter–dominated epoch, such as DESI which has most of its data from that era, primarily constrain the parameter \( c' \), while low-redshift measurements, such as supernovae datasets that are most sensitive to the dark energy–dominated era, serve as probes of \( c \). These two types of observations are nearly independent. For instance, excluding the supernovae datasets leads to the same best-fit value of \( c' \) with \( c\) compatible with $0$ (see Table~\ref{sigmas}), whereas removing DESI and including only supernovae data selects the same value of \( c \) but yields a different best-fit value of \( c'\) compatible with $0$ (see Table~\ref{tab: sigmas without desi}). Moreover, the significances of the individual fits approximately add up to the significance of the combined fit in Table~\ref{sigmas}. This indicates that interpreting \( c \) and \( c' \) as independent variables constrained by independent datasets is a good approximation.

One might wonder how much of our analysis is sensitive to the shape of the scalar potential. For example, let us consider replacing the scalar potential with a hilltop potential in the Fading Dark Sector model. Since DESI prefers a non-zero value of \( c' \), the scalar field rolls significantly during the matter-dominated era regardless of the shape of the scalar potential. Therefore, unless we unnaturally fine-tune the initial condition for $\phi$ such that the scalar field lands at the top of the hilltop at the end of the matter dominated era, we are only sensitive to the slope of the scalar potential in the dark energy dominated era. In particular, as seen in Fig.~\ref{fig:hilltop}, if we choose the initial value of the scalar field to be at the top of the hilltop, it rolls off the hilltop before dark energy dominates, leading to the same picture with an exponential potential with $c>0$.  This agreement confirms that only the local slope of the potential captured by $c$ is sufficient to parameterize $V$ as we have assumed. Adding spatial curvature to the Fading Dark Sector does not improve the fit either as data prefers no spatial curvature (Figs.~\ref{fig:FDS1 and curvature}, \ref{fig:FDS2 and curvature}). Alternatively, if one were to replace the fading dark matter with spatial curvature (as proposed in \cite{Akrami:2025zlb}) the result would be  statistically disfavored relative to the Fading Dark Sector (Table~\ref{tab:sigmas for alternative models}). Our results differ from those in \cite{Akrami:2025zlb} for runs with $\Omega_k\neq0$ due to our use of the full CMB dataset and inclusion of SN samples beyond DESY5, which notably worsen the fits.

Given that the scalar field $\phi$ is very light, with Hubble mass, one might worry about the implications of a fifth force mediated by $\phi$. Let us make two important remarks. First, since visible matter is localized on the Standard Model brane in the extra dimension, it does not couple directly to $\phi$. As a result, no fifth force is mediated within the visible sector. Second, regarding the dark sector, bounds on fifth forces between dark matter particles that rely on standard cosmological evolution do not directly apply to our model, which exhibits a different cosmological history. There are also bounds on fifth forces from galactic-scale observations~\cite{Kesden:2006zb} that must be taken into account. 

The exchange of effectively massless $\phi$ leads to a potential between dark matter constituents.   It is easy to check that in our case this is given in reduced Planck units by
\begin{align}\label{rff}
    \frac{F_{\phi}}{F_{\rm grav}} = 2c'^2\simeq 0.005\pm0.002\,.
\end{align}
The fifth force in the dark sector is strongly constrained to be small. The smallness of the fifth dark force is naturally explained in our model given that $c'$ is $\mathcal{O}(1)$ according to~\eqref{rff}, thereby resolving this apparent fine-tuning problem. In particular, the bound derived in~\cite{Kesden:2006zb} from tidal disruption on dwarf galaxies orbiting the Milky Way suggests $c' \lesssim 0.2$. It is reassuring that the best-fit value from our model, $c' \simeq 0.05\pm0.01$, corresponding to an attractive fifth force in the dark sector $0.5\%$ of the strength of the gravitational force, satisfies the current upper bound. The fact that the upper bound is only slightly above the observed value of $c'$ in our model raises the exciting prospect of detecting a dark fifth force in the near future. It would be interesting and worthwhile to revisit the bounds on the fifth force within our dynamical DM/DE framework.

Our best-fit value for the fifth-force ratio, \( F_\phi/F_{\rm grav} \simeq 0.005 \), is consistent with the analysis of \cite{Bottaro:2024pcb}, which independently found a hint for a 5-th force in the dark sector \( F_5/F_{\rm grav} \simeq 0.004 \pm 0.002 \) in the DESI DR1 data\footnote{We thank Mikhail Ivanov and Tracy Slatyer for bringing this paper to our attention after the first version of this paper appeared.}.

\section{Note added}

The recently recalibrated supernovae datasets, namely DES-Dovekie~\cite{DES:2025sig} and Union3.1/Unity1.8 along with the recalibration of the Pantheon+ dataset~\cite{Hoyt:2026fve,Rubin:2026qdt}, has affected the statistical preference of cosmological models deviating from $\Lambda$CDM. A more consistent significance emerges - the significance of the CPL parameterization reported in the literature appears to be consistently around 3$\sigma$ when fitted to CMB (Planck + ACT), BAO (DESI DR2), and SN (DES-Dovekie, Union3.1, Pantheon+ (recal.)). We similarly observe such a consistent pattern of significance for the fading dark sector model which has now been reported in the newly updated Tab.~\ref{sigmas new}. Furthermore, the recalibrated datasets show a remarkable agreement on the preferred value of $c$ as can be seen in the updated Fig.~\ref{Contours new} for our preferred model with $c<0,c'>0$. This consistency among datasets using supernovae is also observable when using the CPL parameterization. However, the tension between results with and without supernovae datasets for the CPL parameterization remains as can be seen in Fig.~\ref{fig:w0wa contour new}.
In a sense our model could have been used to argue that there is a slight discrepancy between different SN datasets which is now removed after recalibration!
As the measurements of supernovae probe the physics in the energy-dominated era ($z\lesssim 0.5$), this consistency is expected. With this, it is reassuring that the two extended physically motivated parameters in our model (compared to $\Lambda$CDM) are well-measured across all datasets.

\section*{Acknowledgments}
We have greatly benefited from discussions with Cristhian Garcia Quintero and Mustapha Ishak-Boushaki regarding the DESI results.
We would also like to thank Luis Anchordoqui, Daniel Eisenstein, Sonia Paban, and Paul Steinhardt for valuable discussions.

AB is supported in part by the Simons Foundation grant number 654561 and by the Princeton Gravity Initiative at Princeton University. The work of CV and DW is supported in part by a grant from the Simons Foundation (602883, CV) and the DellaPietra Foundation. The work of GO is supported by a
Leverhulme Trust International Professorship grant number LIP-202-014.

\bibliography{main}

\onecolumngrid
\appendix
\section*{Supplemental Figures and Tables}
\label{sec:supplemental}
\setcounter{section}{0}
\renewcommand{\thesection}{S\arabic{section}}

\renewcommand{\thefigure}{S\arabic{figure}}
\setcounter{figure}{0}

\begin{figure}
    \centering
    \includegraphics[width=\linewidth]{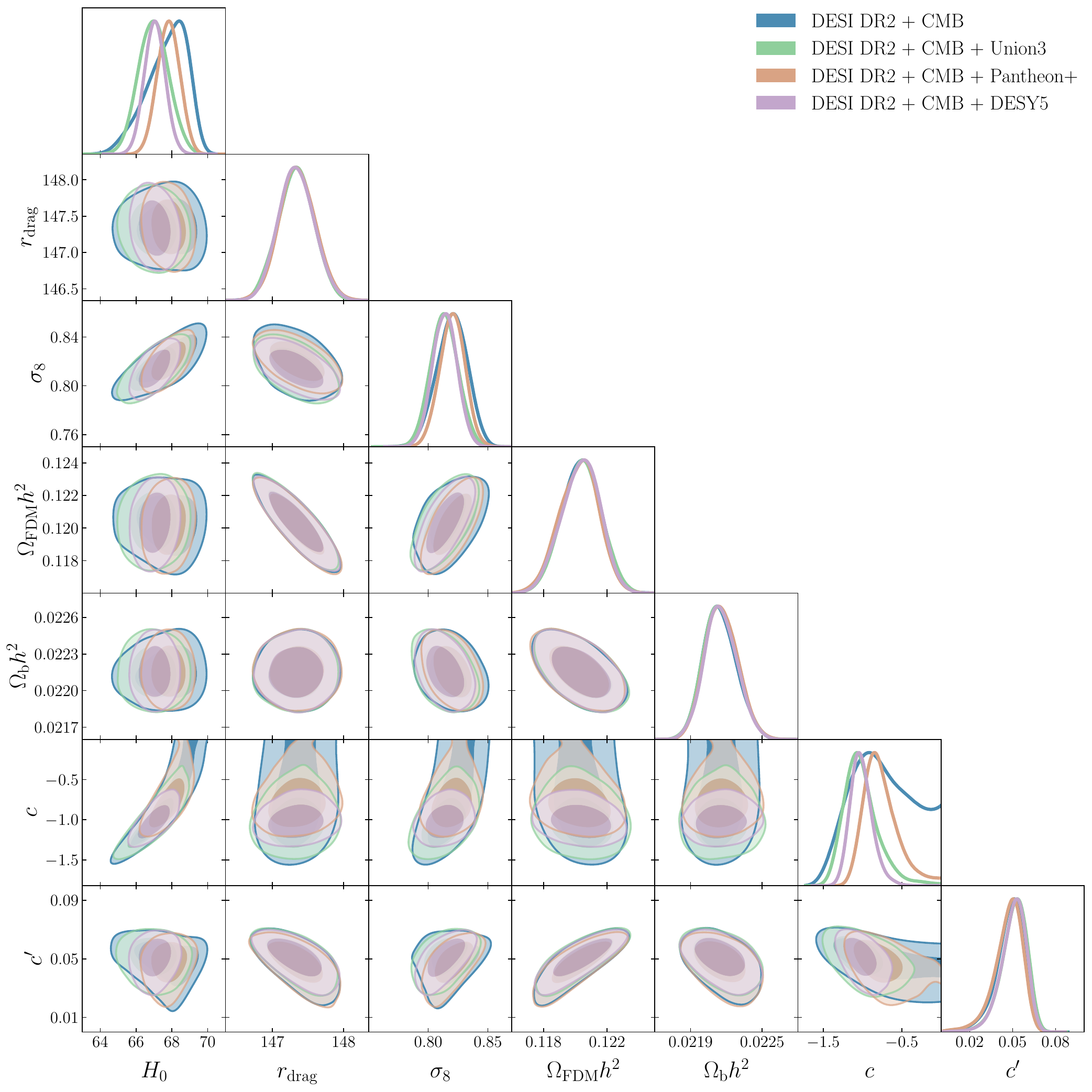}
    \begin{picture}(0,0)\vspace*{-1.2cm}
            \put(-208,0){$[\rm km/s/Mpc]$}
            \put(-125,0){$[\rm Mpc]$}
            \put(-255,402){\rotatebox{90}{$[\rm Mpc]$}}
        \end{picture}
        \vspace{0.5cm}
    \caption{The posterior distributions of the cosmological parameters for the $c\leq 0,c'\geq 0$ model using various datasets.}
    \label{fig:cpcpn}
\end{figure}

\begin{figure}
    \centering
    \includegraphics[width=\linewidth]{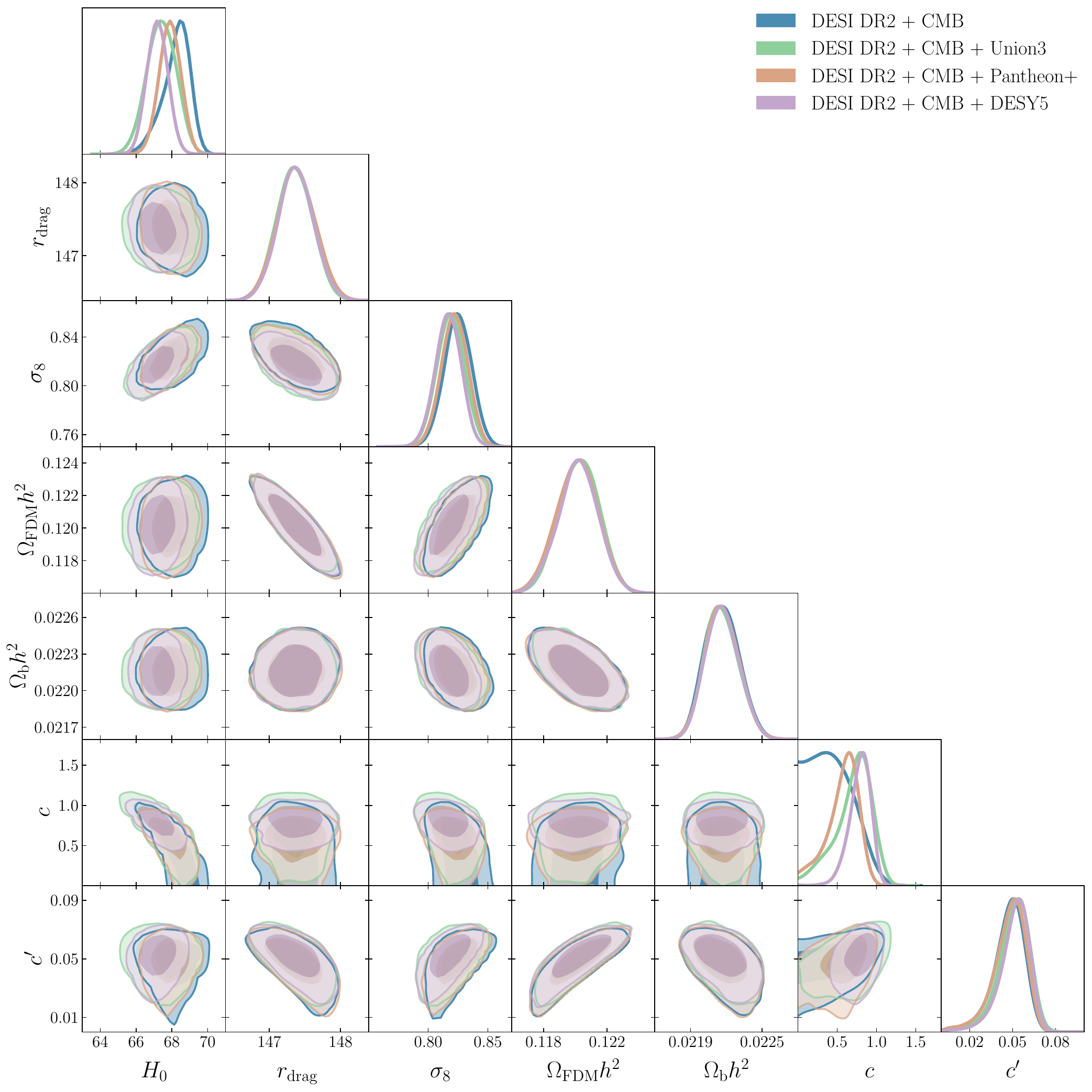}
    \begin{picture}(0,0)\vspace*{-1.2cm}
            \put(-208,0){$[\rm km/s/Mpc]$}
            \put(-125,0){$[\rm Mpc]$}
            \put(-255,402){\rotatebox{90}{$[\rm Mpc]$}}
        \end{picture}
        \vspace{0.5cm}
    \caption{The posterior distributions of the cosmological parameters for the $c,c'\geq 0$ model using various datasets.}
    \label{fig:cpcpp}
\end{figure}

\begin{figure}
    \centering
    \includegraphics[width=\linewidth]{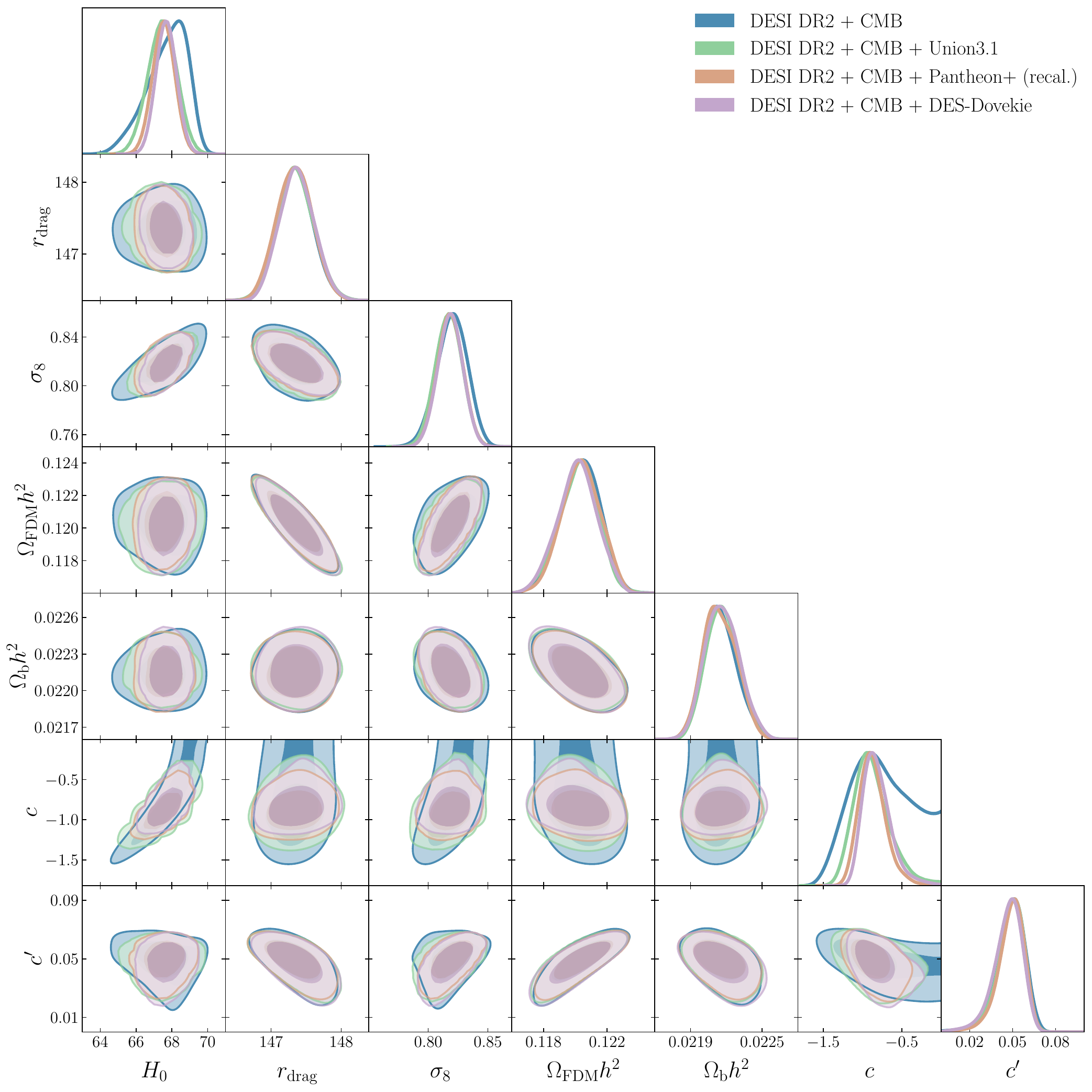}
    \begin{picture}(0,0)\vspace*{-1.2cm}
            \put(-208,0){$[\rm km/s/Mpc]$}
            \put(-125,0){$[\rm Mpc]$}
            \put(-255,402){\rotatebox{90}{$[\rm Mpc]$}}
        \end{picture}
        \vspace{0.5cm}
    \caption{The posterior distributions of the cosmological parameters for the $c\leq 0,c'\geq 0$ model with the recalibrated supernovae data.}
    \label{fig:cpcpn_recal}
\end{figure}

\begin{figure}
    \centering
    \includegraphics[width=\linewidth]{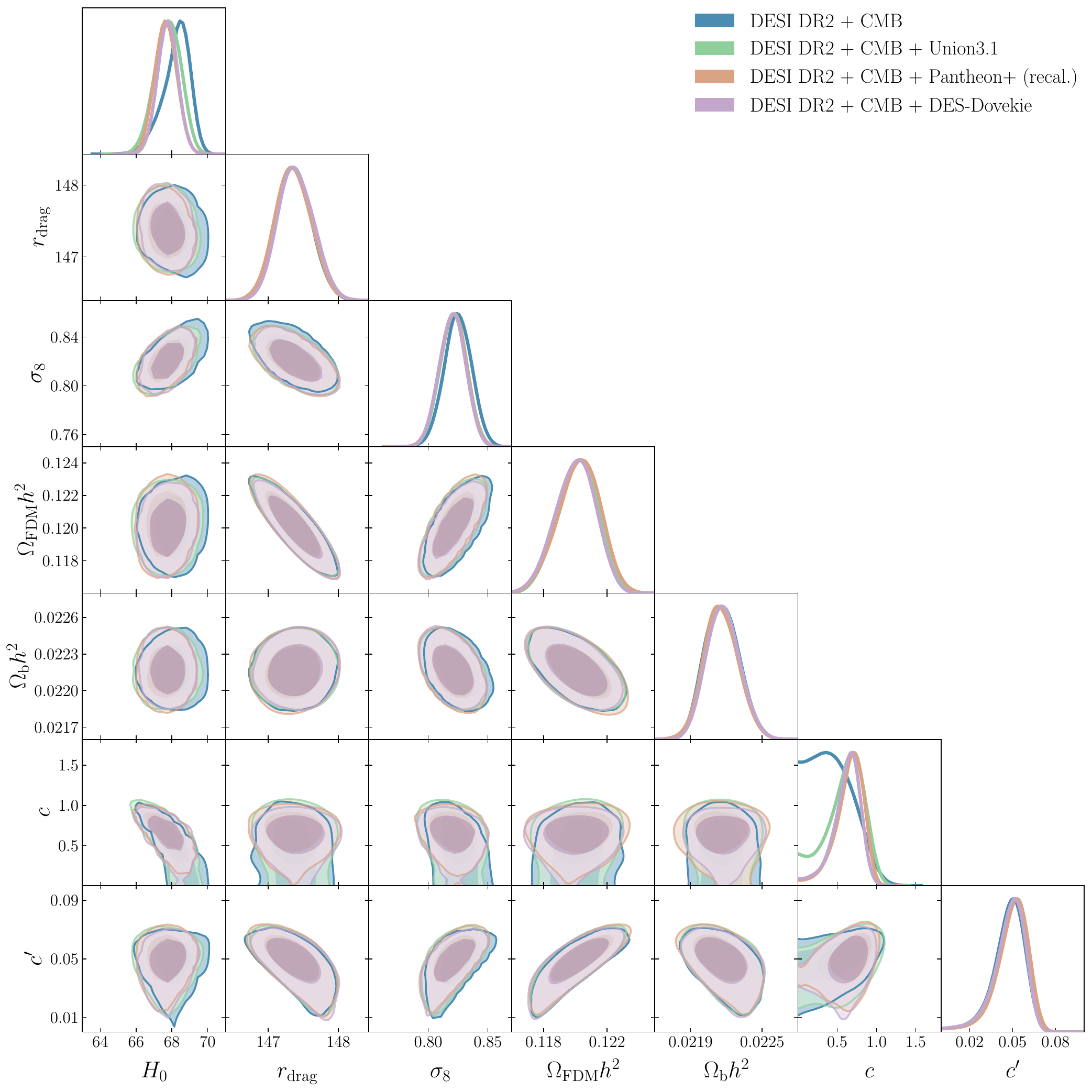}
    \begin{picture}(0,0)\vspace*{-1.2cm}
            \put(-208,0){$[\rm km/s/Mpc]$}
            \put(-125,0){$[\rm Mpc]$}
            \put(-255,402){\rotatebox{90}{$[\rm Mpc]$}}
        \end{picture}
        \vspace{0.5cm}
    \caption{The posterior distributions of the cosmological parameters for the $c,c'\geq 0$ model with the recalibrated supernovae data.}
    \label{fig:cpcpp_recal}
\end{figure}

\begin{figure*}[!tp]
    \centering
    \begin{subfigure}{\textwidth}
        \includegraphics[width=.49\linewidth]{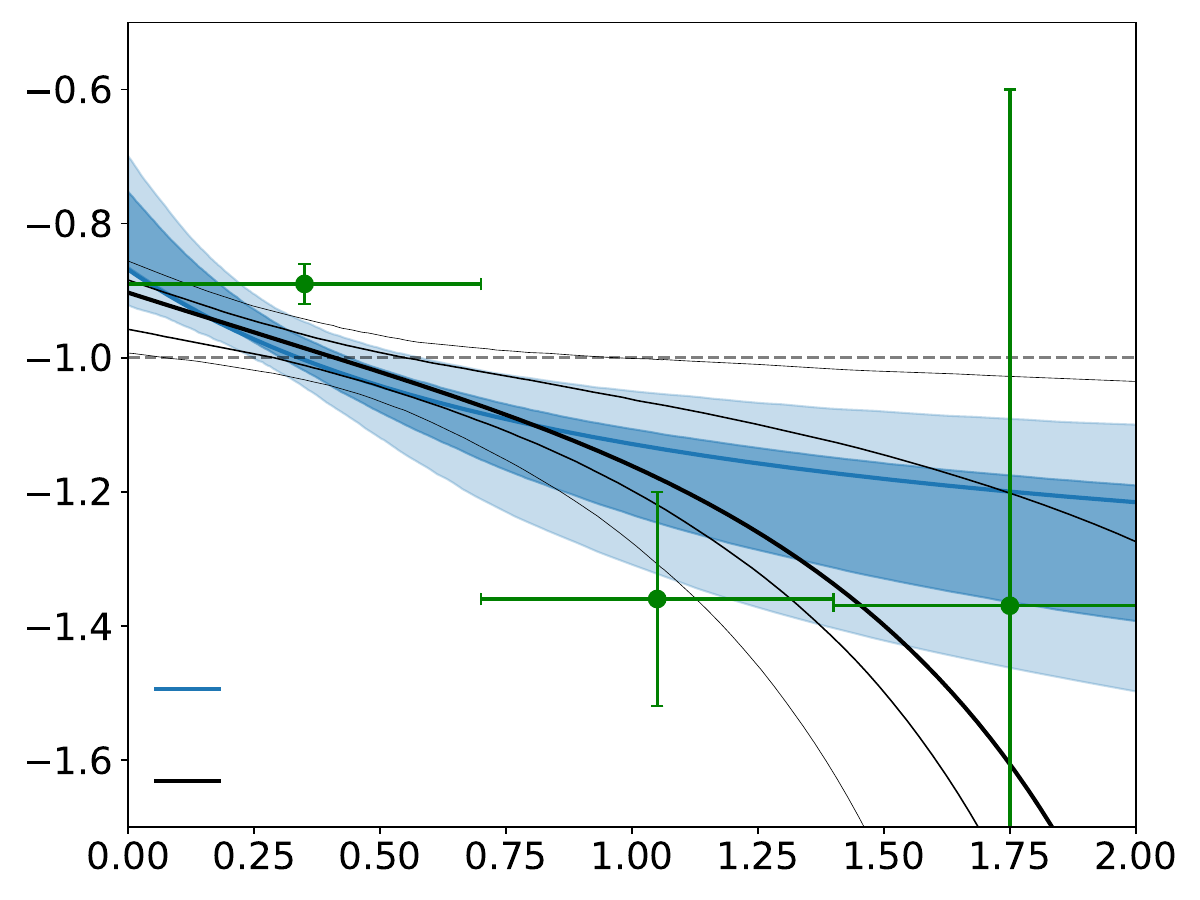}
        \hfill
        \includegraphics[width=.49\linewidth]{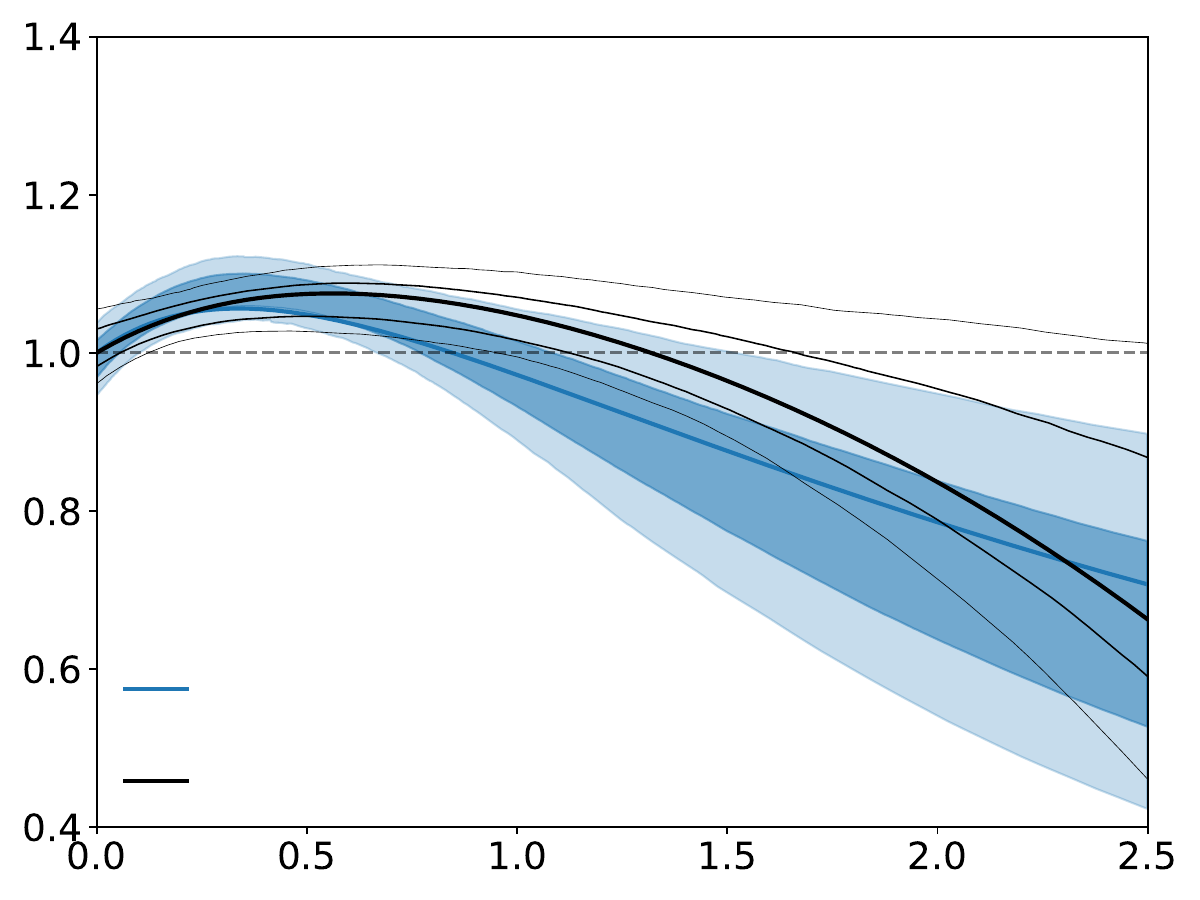}
        \begin{picture}(0,0)\vspace*{-1.2cm}
            \put(-515,80){\rotatebox{90}{$w_{\rm DE,eff}$}}
            \put(-380,-5){$z$}
            \put(-265,80){\rotatebox{90}{$f_{\rm DE,eff}(z)$}}
            \put(-125,-5){$z$}
            \put(-210,41){CPL}
            \put(-210,23){FDS}
            \put(-460,41){CPL}
            \put(-460,23){FDS}
        \end{picture}\vspace*{0.25cm}

        \caption{$c<0,c'>0$}
    \end{subfigure}
    \hfill
    \begin{subfigure}{\textwidth}
        \includegraphics[width=.49\linewidth]{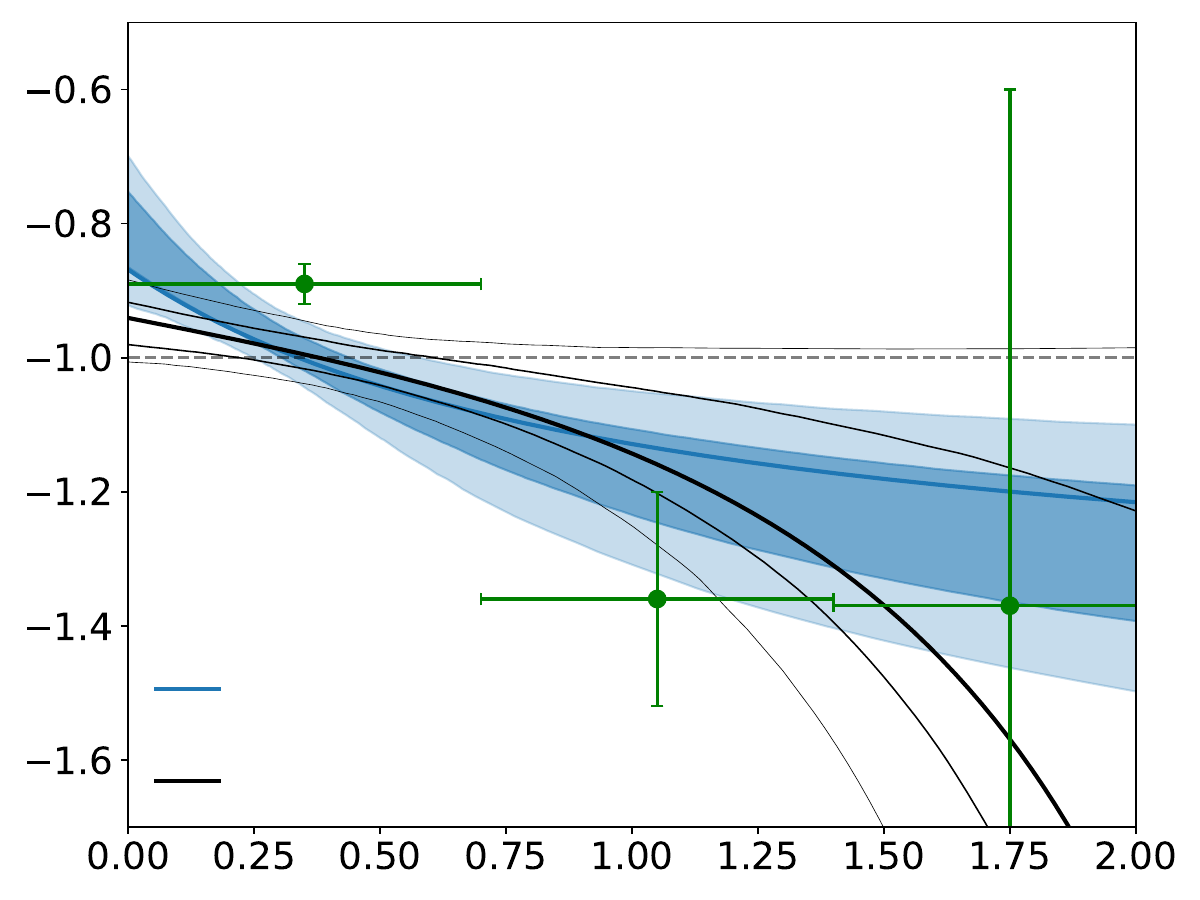}
        \hfill
        \includegraphics[width=.49\linewidth]{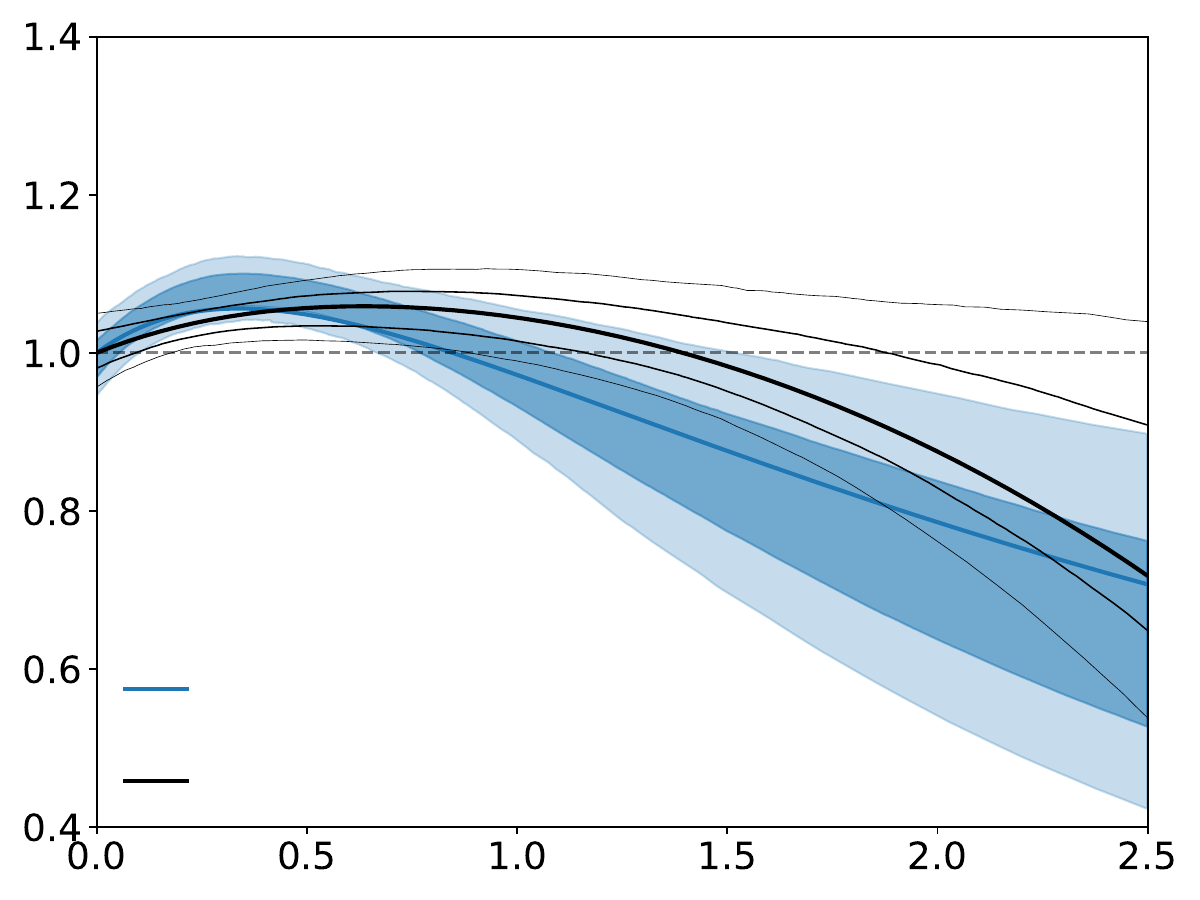}
        \begin{picture}(0,0)\vspace*{-1.2cm}
            \put(-515,80){\rotatebox{90}{$w_{\rm DE,eff}$}}
            \put(-380,-5){$z$}
            \put(-265,80){\rotatebox{90}{$f_{\rm DE,eff}(z)$}}
            \put(-125,-5){$z$}
            \put(-210,41){CPL}
            \put(-210,23){FDS}
            \put(-460,41){CPL}
            \put(-460,23){FDS}
        \end{picture}\vspace*{0.25cm}

        \caption{$c,c'>0$}
    \end{subfigure}
    \caption{An updated version of Fig.~\ref{wplot} using MCMC results fitted to CMB + DESI DR2 + DES-dovekie.}
    \label{fig:updated weff}
\end{figure*}

\begin{figure*}[hp!]
    \centering
    \begin{subfigure}{.49\textwidth}
        \includegraphics[width=\linewidth]{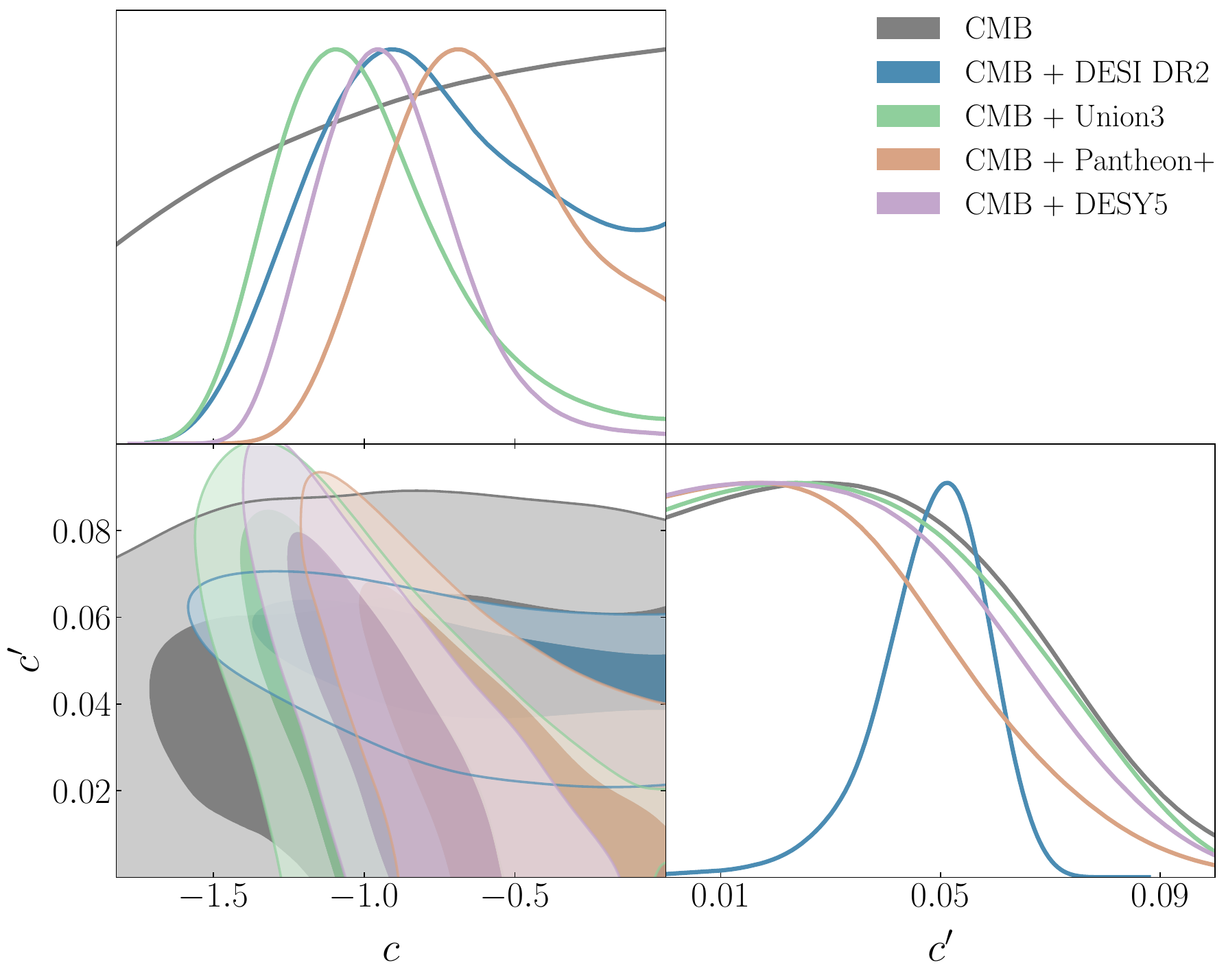}
        \caption{$c<0,c'>0$}
    \end{subfigure}
    \hfill
    \begin{subfigure}{.49\textwidth}
        \includegraphics[width=\linewidth]{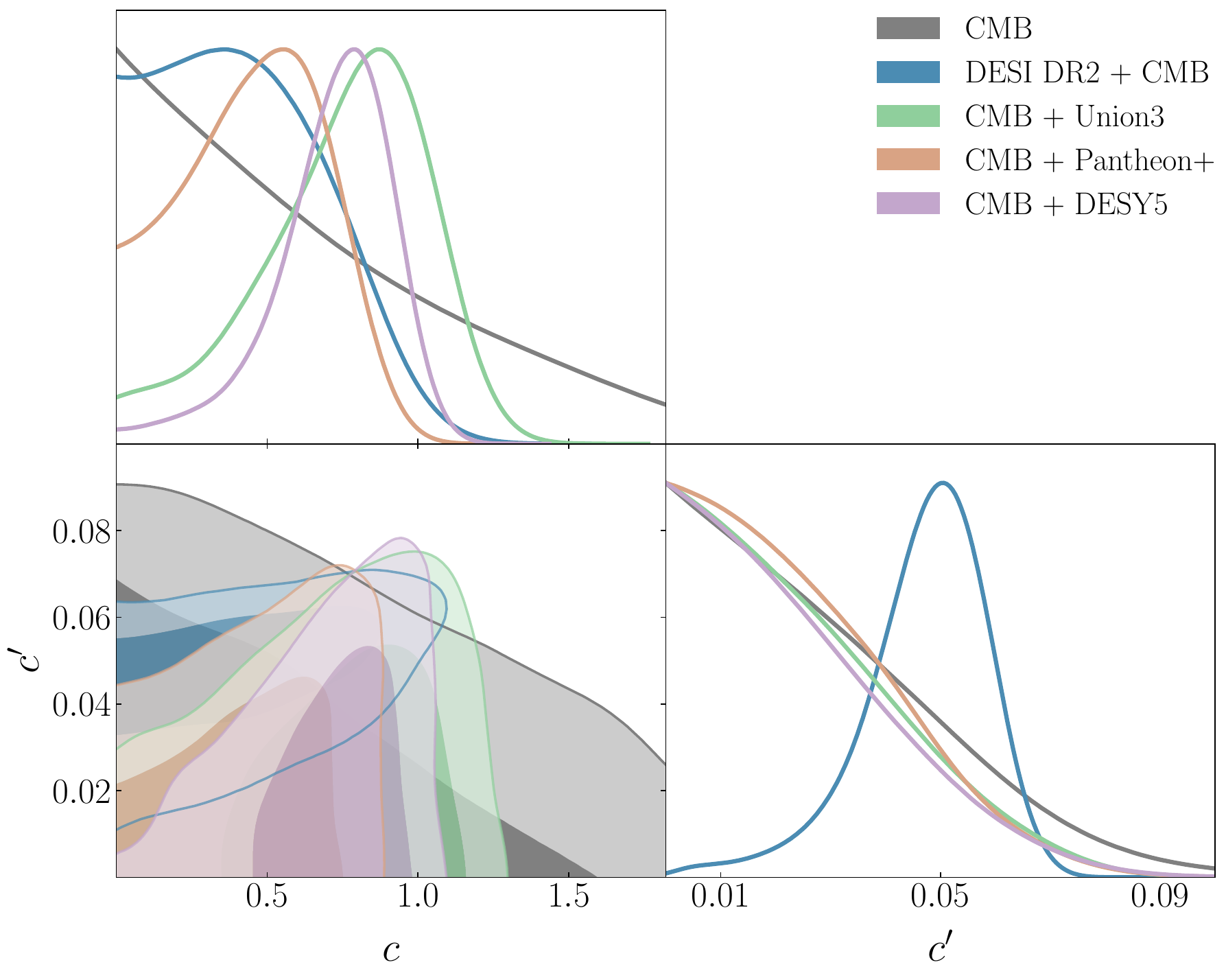}
        \caption{$c>0,c'>0$}
    \end{subfigure}
 \caption{Contours showing the 1- and 2-$\sigma$ constraints on $c$ and $c'$, along with their marginalized posteriors for two fading dark sector models with opposite signs of $c$, across various datasets. $\Lambda$CDM corresponds to $c=c'=0$.}
\label{fig: FDS - without DESI DR2 + SN Contours}
\end{figure*}

\begin{figure}
    \centering
    \includegraphics[width=\linewidth]{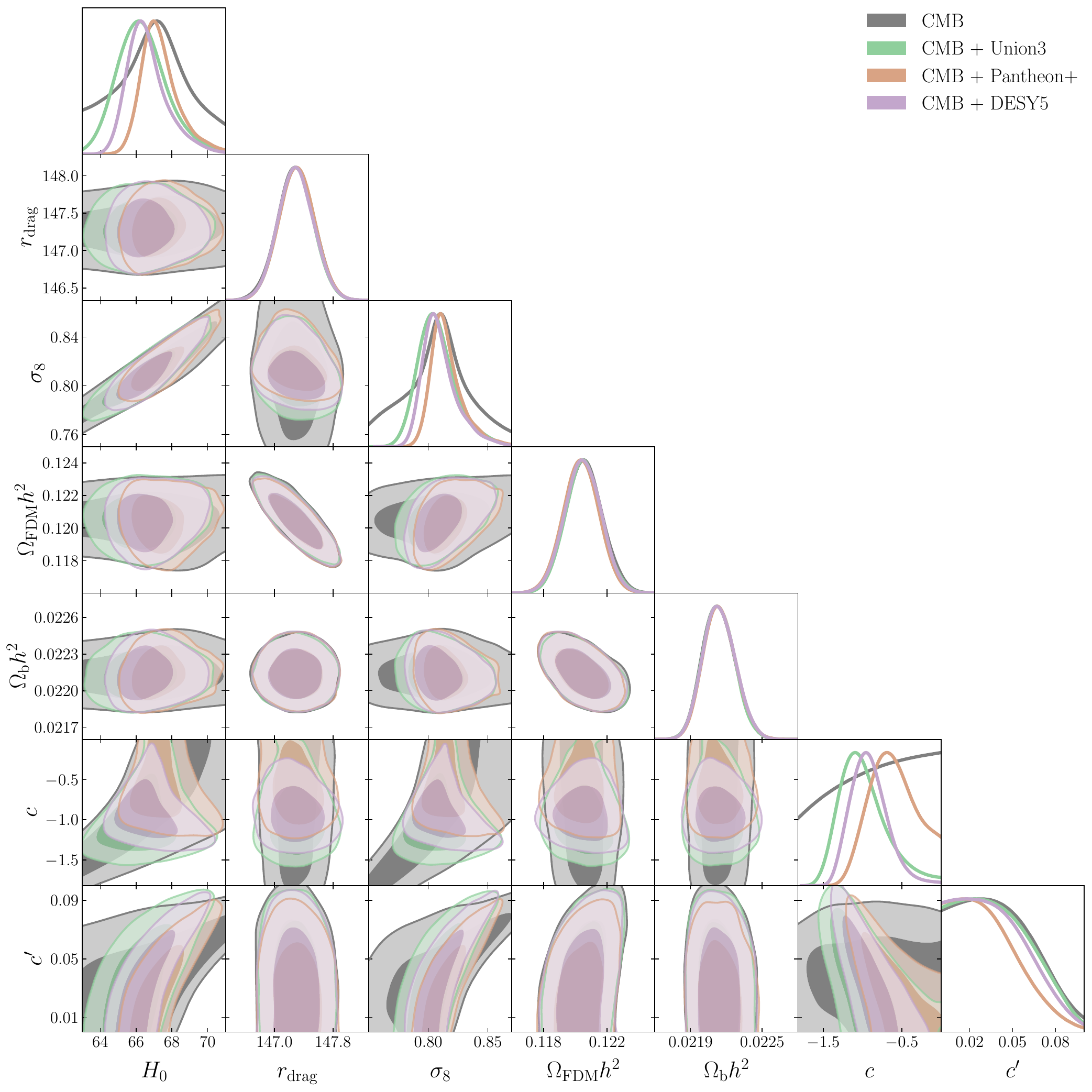}
    \begin{picture}(0,0)\vspace*{-1.2cm}
            \put(-208,0){$[\rm km/s/Mpc]$}
            \put(-125,0){$[\rm Mpc]$}
            \put(-255,402){\rotatebox{90}{$[\rm Mpc]$}}
        \end{picture}
        \vspace{0.5cm}
    \caption{The posterior distributions of the cosmological parameters for the $c\leq 0,c'\geq 0$ model using various datasets, \textit{excluding DESI DR2}.}
    \label{fig:cpcpn CMB}
\end{figure}

\begin{figure}
    \centering
    \includegraphics[width=\linewidth]{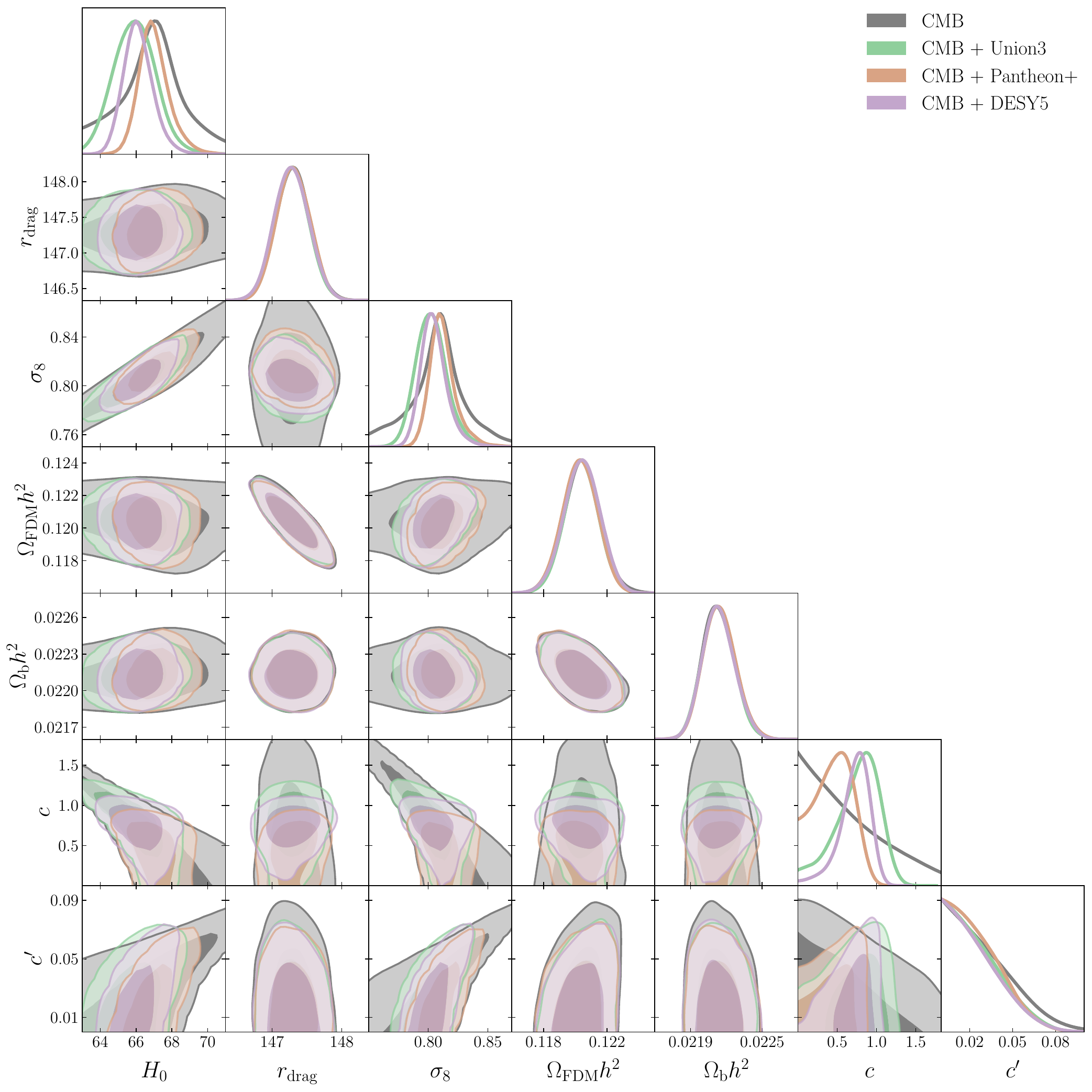}
    \begin{picture}(0,0)\vspace*{-1.2cm}
            \put(-208,0){$[\rm km/s/Mpc]$}
            \put(-125,0){$[\rm Mpc]$}
            \put(-255,402){\rotatebox{90}{$[\rm Mpc]$}}
        \end{picture}
        \vspace{0.5cm}
    \caption{The posterior distributions of the cosmological parameters for the $c\geq 0,c'\geq 0$ model using various datasets, \textit{excluding DESI DR2}.}
    \label{fig:cpcpp CMB}
\end{figure}

\begin{figure}
    \centering
    \includegraphics[width=\linewidth]{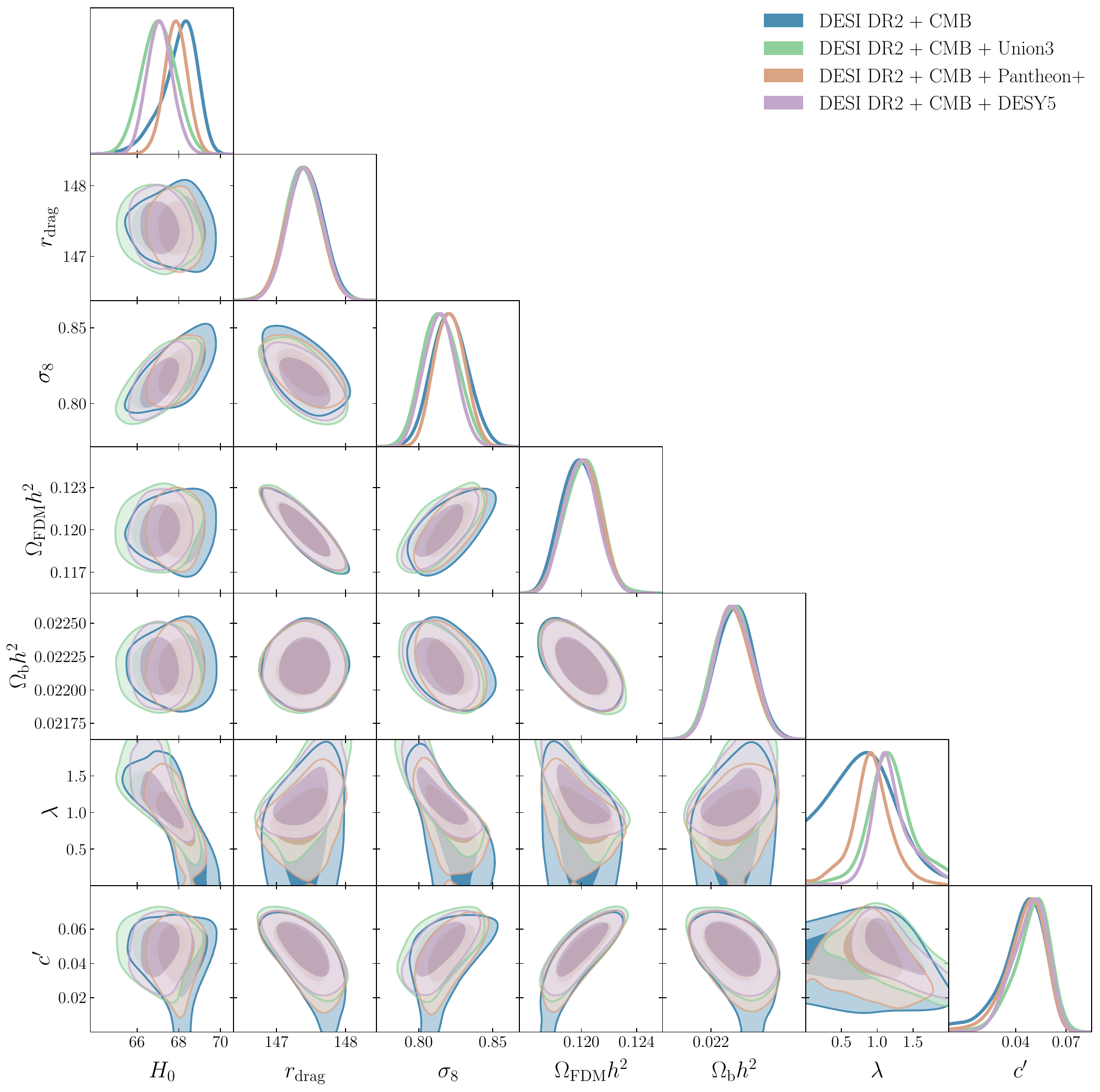}
    \begin{picture}(0,0)\vspace*{-1.2cm}
            \put(-208,0){$[\rm km/s/Mpc]$}
            \put(-125,0){$[\rm Mpc]$}
            \put(-255,402){\rotatebox{90}{$[\rm Mpc]$}}
        \end{picture}
        \vspace{0.5cm}
    \caption{The posterior distributions of the cosmological parameters for the coupled dark sector model with $V=V_0\exp(-\lambda\phi^2)$ and $m=m_0\exp(-c'\phi)$ using various datasets.}
    \label{fig:hilltop}
\end{figure}

\begin{figure}
    \centering
    \includegraphics[width=\linewidth]{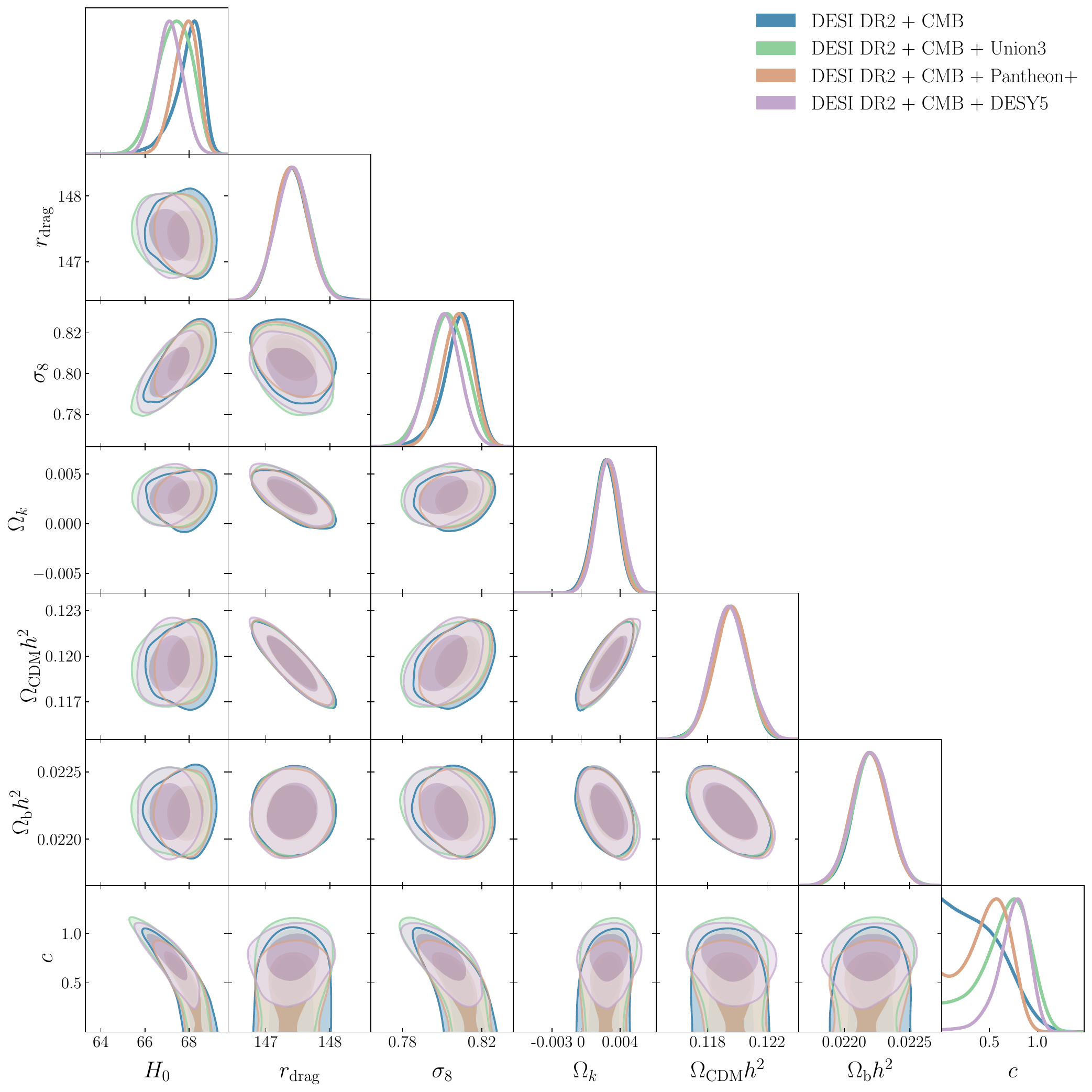}
    \begin{picture}(0,0)\vspace*{-1.2cm}
            \put(-208,0){$[\rm km/s/Mpc]$}
            \put(-125,0){$[\rm Mpc]$}
            \put(-255,402){\rotatebox{90}{$[\rm Mpc]$}}
        \end{picture}
        \vspace{0.5cm}
    \caption{The posterior distributions of the cosmological parameters for the quintessence model with $V=V_0\exp(-c\phi)$ and varying curvature $\Omega_K$ using various datasets. Here, we have restricted to $c>0$. Additionally, the best-fit values of $\Omega_K$ among the above datasets are $0.002$ without supernovae and $0.003$ with supernovae.}
    \label{fig:quintessence and curvature}
\end{figure}

\begin{figure}
    \centering
    \includegraphics[width=\linewidth]{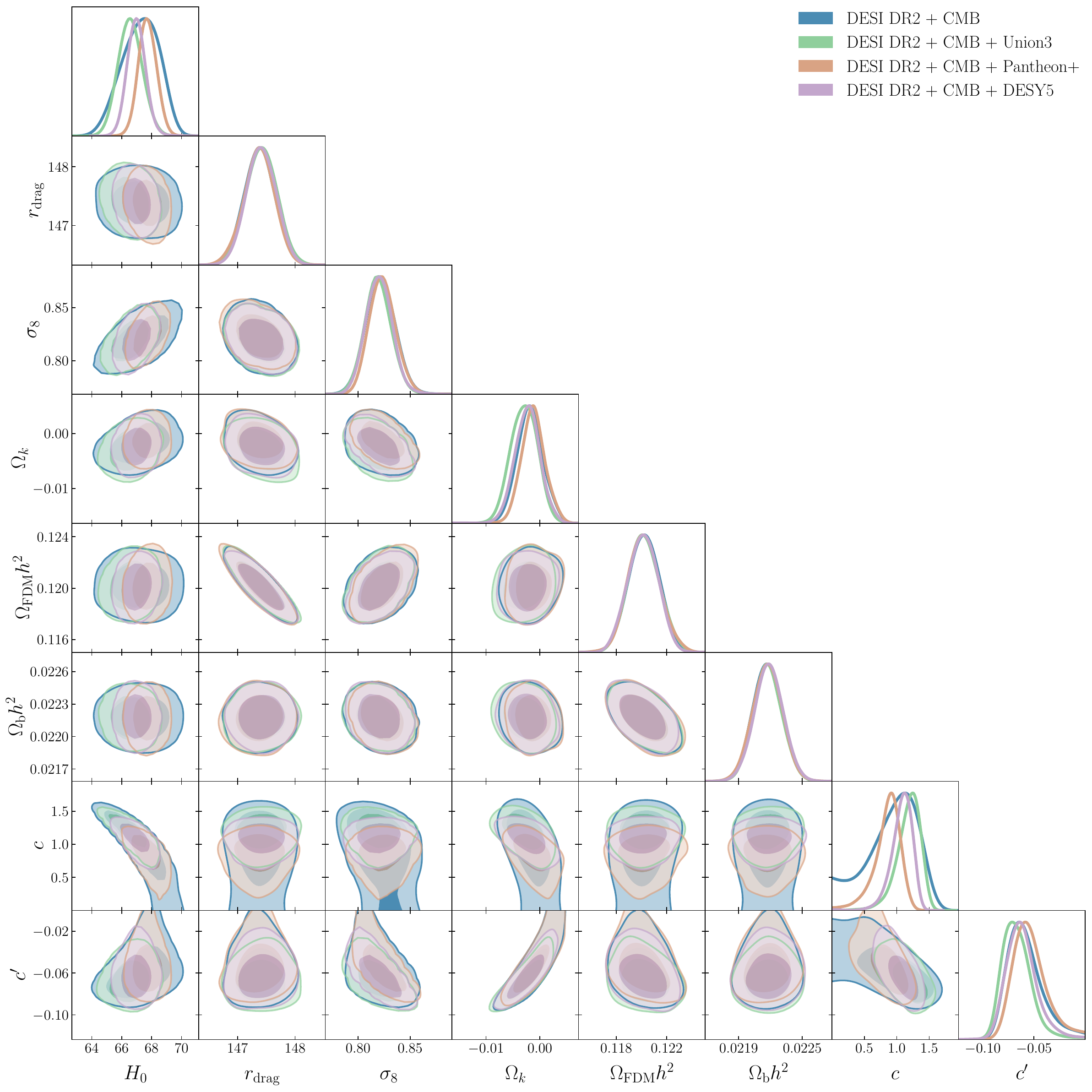}
    \begin{picture}(0,0)\vspace*{-1.2cm}
            \put(-215,0){$[\rm km/s/Mpc]$}
            \put(-142,0){$[\rm Mpc]$}
            \put(-255,417){\rotatebox{90}{$[\rm Mpc]$}}
        \end{picture}
        \vspace{0.5cm}
    \caption{The posterior distributions of the cosmological parameters for the FDS model with $c\leq 0,c'\geq 0$ while allowing variation in the curvature $\Omega_K$ using various datasets.}
    \label{fig:FDS1 and curvature}
\end{figure}

\begin{figure}
    \centering
    \includegraphics[width=\linewidth]{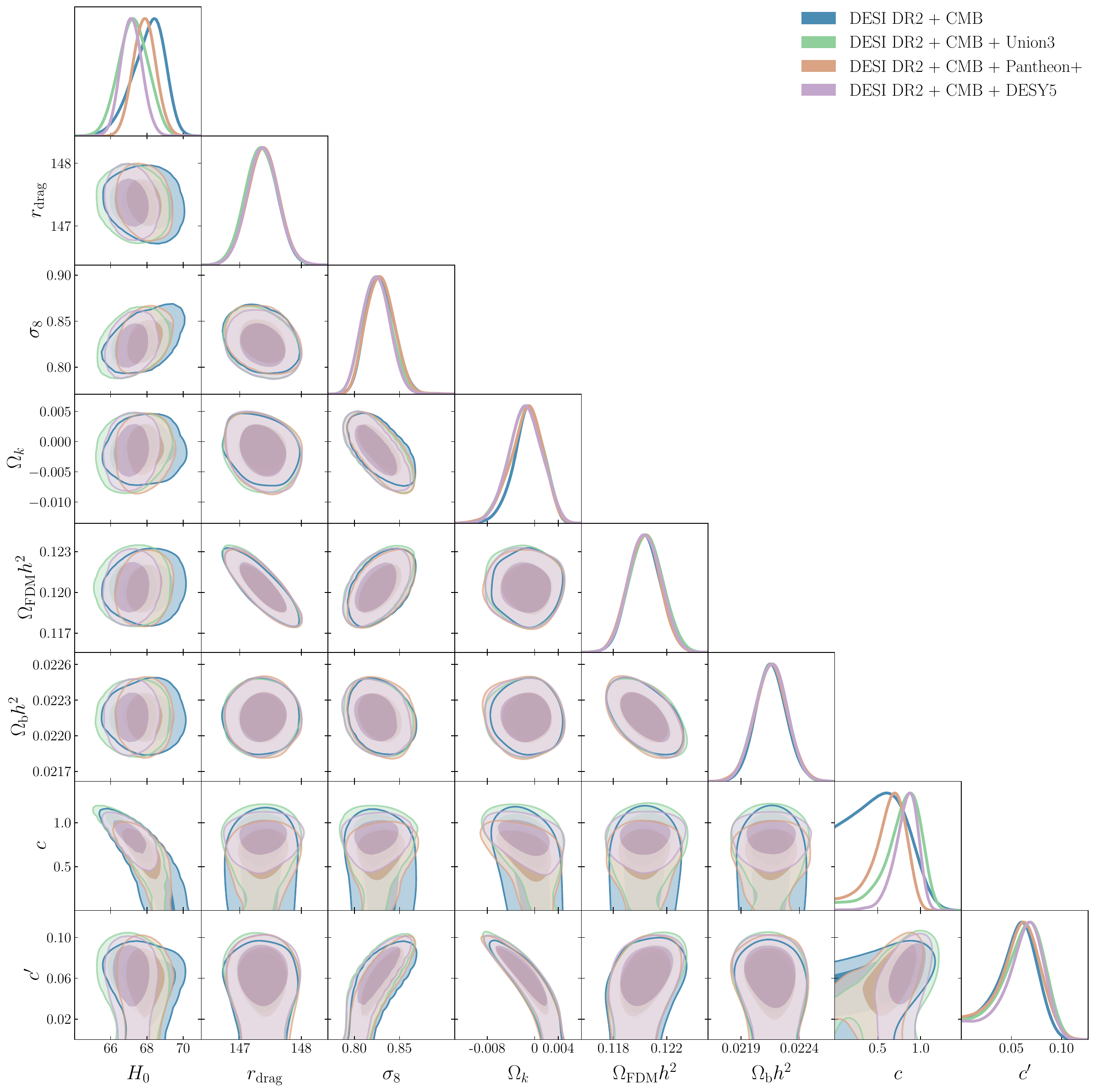}
    \begin{picture}(0,0)\vspace*{-1.2cm}
            \put(-215,0){$[\rm km/s/Mpc]$}
            \put(-142,0){$[\rm Mpc]$}
            \put(-255,417){\rotatebox{90}{$[\rm Mpc]$}}
        \end{picture}
        \vspace{0.5cm}
    \caption{The posterior distributions of the cosmological parameters for the FDS model with $c,c'\geq 0$ while allowing variation in the curvature $\Omega_K$ using various datasets.}
    \label{fig:FDS2 and curvature}
\end{figure}

\begin{figure}
    \centering
    \includegraphics[width=\linewidth]{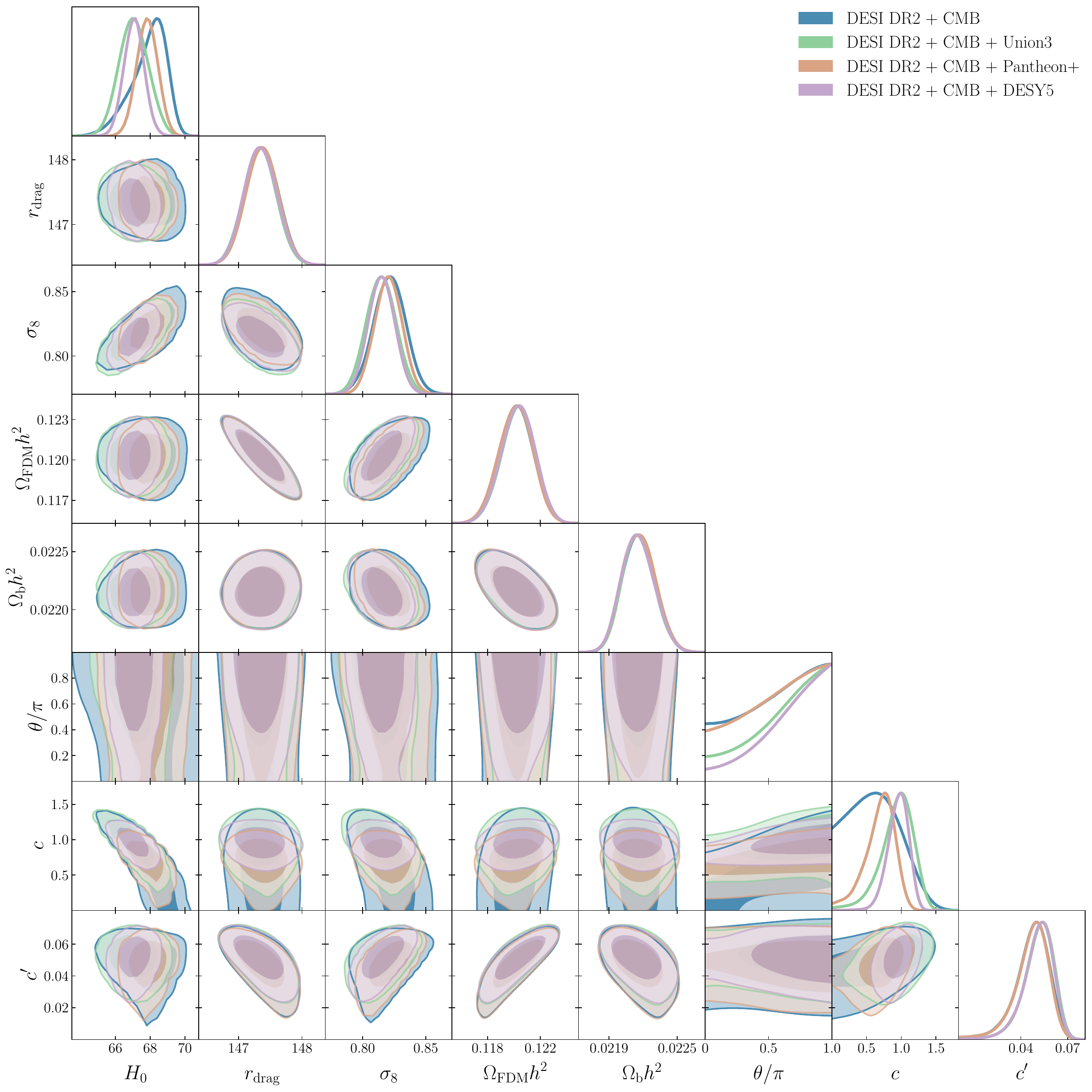}
    \begin{picture}(0,0)\vspace*{-1.2cm}
            \put(-215,0){$[\rm km/s/Mpc]$}
            \put(-142,0){$[\rm Mpc]$}
            \put(-255,417){\rotatebox{90}{$[\rm Mpc]$}}
        \end{picture}
        \vspace{0.5cm}
    \caption{The posterior distributions of the cosmological parameters for the FDS model with two light scalar fields with $c,c'\geq 0$. Here, we have $V=V_0\exp(-c\phi)$ and $m=m_0\exp(-c'[\phi_1\cos\theta+\phi_2\sin\theta])$ where $\theta\in[0,\pi]$}
    \label{fig:FDS 2 scalar fields}
\end{figure}

\begin{table*}[ht!]
\centering
\begin{tabular}{|c|c|>{\hspace{.2cm}}c<{\hspace{.2cm}}|c|c|c|>{\hspace{.2cm}}c<{\hspace{.2cm}}|c|c|}
\hline
\multirow{2}{*}{Datasets} &  \multicolumn{4}{c|}{FDS ($c<0$)}& \multicolumn{4}{c|}{FDS ($c>0$)}\\
\cline{2-9}
& $\Delta\chi^2_{\rm MAP}$ & $\sigma$ & $c$ & $c'$ & $\Delta\chi^2_{\rm MAP}$ & $\sigma$ & $c$ & $c'$\\
\hline
\multirow{2}{*}{CMB} &  \multirow{2}{*}{$0.0$} & \multirow{2}{*}{0.0} & $-0.22$ & $0.01$ & \multirow{2}{*}{$0.1$} & \multirow{2}{*}{0.0} & $0.29$& $0.00$\\
&  &  & $-0.88^{+0.62}_{-0.65}$ & $0.04\pm 0.03$ & & & $0.64^{+0.55}_{-0.50}$ & $0.03\pm 0.02$\\ 
\arrayrulecolor{gray!30}\hline
\multirow{2}{*}{CMB+Union3} & \multirow{2}{*}{$-3.3$} & \multirow{2}{*}{$1.3$} & $-1.07$ & $0.04$ & \multirow{2}{*}{$-2.8$} & \multirow{2}{*}{$1.2$} & $0.85$ & $0.00$\\
 &  &  & $-0.99\pm 0.30$ & $0.04\pm 0.03$ & & & $0.76^{+0.27}_{-0.28}$ & $0.03\pm 0.02$\\ 
 \arrayrulecolor{gray!30}\hline
\multirow{2}{*}{CMB+Pantheon+}  & \multirow{2}{*}{$-0.7$} & \multirow{2}{*}{$0.4$} & $-0.59$ & $0.02$ & \multirow{2}{*}{$-0.6$} & \multirow{2}{*}{$0.3$} & $0.46$ & $0.00$\\
 &  &  & $-0.61^{+0.33}_{-0.30}$ & $0.03\pm 0.02$ & & & $0.46^{+0.25}_{-0.27}$ & $0.03\pm 0.02$\\ 
 \arrayrulecolor{gray!30}\hline
\multirow{2}{*}{CMB+DESY5} &  \multirow{2}{*}{$-5.1$} & \multirow{2}{*}{$1.8$} & $-0.98$ & $0.04$ & \multirow{2}{*}{$-4.7$} & \multirow{2}{*}{$1.7$} & $0.78$ & $0.00$\\
 &  &  & $-1.00^{+0.13}_{-0.14}$ & $0.05\pm 0.01$ & & & $0.72^{+0.19}_{-0.18}$ & $0.02\pm 0.02$\\ 
\arrayrulecolor{black}\hline
\end{tabular}
\caption{Statistical significance comparing the fading dark sector models for both $c > 0$ and $c < 0$ alongside the CPL parametrization, across various combinations of supernova datasets, to $\Lambda$CDM. Additionally, the best fit, the mean, and the $\pm 1\sigma$ values for $c,c'$ are provided for each fading dark sector model.}
\label{tab: sigmas without desi}
\end{table*}

\begin{table*}[ht!]
\centering
\begin{tabular}{|c|c|>{\hspace{.5cm}}c<{\hspace{.5cm}}|c|>{\hspace{.2cm}}c<{\hspace{.2cm}}|c|c|c|>{\hspace{.2cm}}c<{\hspace{.2cm}}|}
\hline
\multirow{2}{*}{Datasets} & \multicolumn{2}{c|}{Hilltop FDS} & \multicolumn{2}{c|}{Quintessence + $\Omega_K$}& \multicolumn{2}{c|}{FDS + $\Omega_K$ ($c<0$)}&
\multicolumn{2}{c|}{FDS + $\Omega_K$ ($c>0$)}\\
\cline{2-9}
& $\Delta\chi^2_{\rm MAP}$ & $\sigma$ & $\Delta\chi^2_{\rm MAP}$ & $\sigma$ & $\Delta\chi^2_{\rm MAP}$ & $\sigma$ & $\Delta\chi^2_{\rm MAP}$ & $\sigma$ \\
\hline
DESI+CMB & $-6.7$ & $2.1$ & $-4.0$ & $1.5$ & $-10.1$ & $2.4$ & $-7.5$ & $1.9$\\ 
\arrayrulecolor{gray!30}\hline
DESI+CMB+Union3 & $-11.8$ & $3.0$ & $-7.1$ & $2.2$ & $-16.6$ & $3.3$ & $-12.0$ & $2.7$ \\ 
 \arrayrulecolor{gray!30}\hline
DESI+CMB+Pantheon+ & $-9.7$ & $2.7$ & $-6.0$ & $2.0$ & $-12.6$ & $2.8$ & $-10.1$ & $2.4$ \\ 
 \arrayrulecolor{gray!30}\hline
DESI+CMB+DESY5
 & $-16.9$ & $3.7$ & $-12.1$ & $3.0$ & $-21.2$ & $3.9$ & $-17.1$ & $3.4$ \\ 
\arrayrulecolor{black}\hline
\end{tabular}
\caption{Statistical significance comparing various alternative models to $\Lambda$CDM.}
\label{tab:sigmas for alternative models}
\end{table*}

\end{document}